\documentclass[12pt,a4paper]{article}
\usepackage{color}
\usepackage{amsmath}
\usepackage{array}
\usepackage[colorlinks=true]{hyperref}
\usepackage[margin=1in]{geometry}
\usepackage{fullpage,mathrsfs,graphicx,framed,color,amssymb,amsmath,amsthm}
\usepackage[boxed, noline, ruled, linesnumbered]{algorithm2e}
\usepackage{epsfig, graphicx}
\usepackage{graphicx}
\usepackage{latexsym,amsfonts,amsbsy,amssymb}
\usepackage{amsmath,amsthm}
\usepackage{enumerate}
\usepackage{cleveref}
\usepackage{changes} 
\makeatletter 

\@addtoreset{equation}{section}
\makeatother 
\newcommand{\R}{\mathbb{R}}

\allowdisplaybreaks 
\newtheorem{theorem}{Theorem}[section]

\allowdisplaybreaks 

\numberwithin{equation}{section}
\begin{document}
\title{Solving Schr\"{o}dinger Equation Using Tensor Neural Network\footnote{This work was 
supported by the National Key Research and Development Program of 
China (2023YFB3309104), National Natural Science Foundations of China (NSFC 1233000214), 
National Key Laboratory of Computational Physics (No. 6142A05230501),  
Beijing Natural Science Foundation (Z200003), National Center for Mathematics 
and Interdisciplinary Science, CAS.}}
\author{Yangfei Liao\footnote{LSEC, NCMIS, Institute
of Computational Mathematics, Academy of Mathematics and Systems
Science, Chinese Academy of Sciences, Beijing 100190,
China,  and School of Mathematical Sciences, University
of Chinese Academy of Sciences, Beijing 100049, China (liaoyangfei@lsec.cc.ac.cn).},\ \ 
Zhongshuo Lin\footnote{LSEC, NCMIS, Institute
of Computational Mathematics, Academy of Mathematics and Systems
Science, Chinese Academy of Sciences, Beijing 100190,
China,  and School of Mathematical Sciences, University
of Chinese Academy of Sciences, Beijing 100049, China (linzhongshuo@lsec.cc.ac.cn).}, \ \ 
Jianghao Liu\footnote{Zhiyuan College, Department of Mathematics 
and Applied Mathematics, Shanghai Jiao Tong University, 
Shanghai 201100, China (jh-liu@sjtu.edu.cn).}, \ \ 
Qingyuan Sun\footnote{Zhiyuan College, Department of Mathematics 
and Applied Mathematics, Shanghai Jiao Tong University, 
Shanghai 201100, China (1025757709@sjtu.edu.cn).}, \\  
Yifan Wang\footnote{School of Mathematical Sciences, Peking University, Beijing 100871, 
China (wangyifan1994@pku.edu.cn).},\ \   
Teng Wu\footnote{Zhiyuan College, Department of Mathematics and Applied Mathematics, 
Shanghai Jiao Tong University, Shanghai 201100, China (sdwt2022sjtu@sjtu.edu.cn).}, \ \ 
Hehu Xie\footnote{LSEC, NCMIS, Institute 
of Computational Mathematics, Academy of Mathematics and Systems
Science, Chinese Academy of Sciences, Beijing 100190,
China,  and School of Mathematical Sciences, University
of Chinese Academy of Sciences, Beijing 100049, China (hhxie@lsec.cc.ac.cn).}
\ \ \ and \ \ 
Mingfeng He\footnote{Zhiyuan College, Department of Mathematics and Applied Mathematics, 
Shanghai Jiao Tong University, Shanghai 201100, China (18666263369@sjtu.edu.cn). 
All contributions by this author involve conducting numerical tests for the interpolation method.}}

\date{}
\maketitle

\begin{abstract}
In this paper, we introduce a novel approach to solve the many-body 
Schr\"{o}dinger equation by the tensor neural network.
Based on the tensor product structure, we can do the direct numerical integration 
by using fixed quadrature points for the functions constructed by the tensor 
neural network within tolerable computational complexity. 
Especially, we design several types of efficient numerical methods to treat 
the variable-coupled Coulomb potentials with high accuracy. 
The corresponding machine learning method is built for solving
many-body Schr\"{o}dinger equation. Some numerical examples are provided to 
validate the accuracy and efficiency of the proposed algorithms.

\end{abstract}

\section{Introduction}
The Schr\"{o}dinger equation is the most fundamental problem in quantum mechanics, 
which is named after Erwin Schr\"{o}dinger, who won the Nobel Prize along with Paul 
Dirac in 1933 for their contributions to quantum physics.
Schr\"{o}dinger equation describes the wave function of a quantum mechanical system, 
which gives probabilistic information about the location of a particle and other 
observable quantities such as its momentum, energy \cite{dirac1929quantum}.
The physical system, and different values for observable quantities can be obtained 
by applying the corresponding operators to the wave functions.

Under the Born-Oppenheimer approximation \cite{born1985quantentheorie}, 
the system of $N$ electrons and $M$ ions is described by the following Hamiltonian
\begin{eqnarray}\label{Hamiltonian_Operator}
\widehat H = -\frac{1}{2}\sum_{i=1}^N \Delta_i
+ \sum_{i=1}^N \sum_{j=i+1}^N \frac{1}{|\mathbf r_i-\mathbf r_j|} 
-\sum_{I=1}^M\sum_{i=1}^N \frac{Z_I}{|\mathbf r_i-\mathbf R_I|}
+\sum_{I=1}^M\sum_{J=I+1}^M\frac{Z_IZ_J}{|\mathbf R_I-\mathbf R_J|},
\end{eqnarray}
where $\mathbf r = (\mathbf r_1, \cdots, \mathbf r_N)$ and 
$\mathbf R=(\mathbf R_1, \cdots, \mathbf R_M)$ denote the coordinates 
of electrons and ions, respectively, $Z_I$ is the nuclear charge for the $I$-th ion.
Since the Hamiltonian (\ref{Hamiltonian_Operator}) is spin-independent, 
the wave function  can be written as the time-independent form 
$\Psi(\mathbf r,\mathbf R)$. For easy description and understanding, 
we concentrate on computing the electronic structure of the ground 
state of the system with the Hamiltonian
\begin{eqnarray}\label{Hamiltonian_Operator_2}
\widehat H = -\frac{1}{2}\sum_{i=1}^N \Delta_i
+ \sum_{i=1}^N \sum_{j=i+1}^N \frac{1}{|\mathbf r_i-\mathbf r_j|} 
-\sum_{I=1}^M\sum_{i=1}^N \frac{Z_I}{|\mathbf r_i-\mathbf R_I|}.
\end{eqnarray}
Then the wave function  can be denoted by $\Psi(\mathbf r)$.
And the ground state of the system can be acquired directly by 
minimizing the following energy functional
\begin{eqnarray}\label{Energy_Definition}
E[\Psi] = \frac{\langle \Psi|\widehat H|\Psi\rangle}{\langle \Psi|\Psi\rangle},
\end{eqnarray}
which will be called the Schr\"{o}dinger equation in this paper.

Unfortunately, there are two main difficulties in solving Schr\"{o}dinger 
equation (\ref{Energy_Definition}). The first difficulty is that 
(\ref{Energy_Definition}) is a high-dimensional optimization problem 
and the dimension of the wave function  $\Psi(\mathbf r)$ is $3N$.
Both the number of grids meshed from the $\mathbf r$ space and the 
computational complexity of direct quadrature scheme grow exponentially in $N$.
This crisis which is known as the curse of dimensionality 
(CoD) \cite{bellman1957dynamic} leads to that (\ref{Energy_Definition}) 
is almost impossible solved by traditional numerical methods.
The second difficulty is that the Pauli exclusion principle must be imposed 
on the wave function  \cite{pauli1925zusammenhang}.
Based on Pauli exclusion principle, the wave function  should satisfy 
the anti-symmetry property, i.e., the following equality holds for
$1\leq i\neq j \leq N$
\begin{eqnarray}\label{Anti_Symmetry}
\frac{\langle T_{ij}\Psi|\Psi\rangle }{\langle\Psi|\Psi\rangle}=-1,
\end{eqnarray}
where $T_{ij}$ denotes the exchange operator for different position of electrons
\begin{eqnarray}\label{Exchange_Operator}
T_{ij}\Psi(\mathbf r_1,\cdots,\mathbf r_i,\cdots,\mathbf r_j,\cdots, \mathbf r_N)
= \Psi(\mathbf r_1,\cdots,\mathbf r_j,\cdots,\mathbf r_i,\cdots, \mathbf r_N), \ \ \
1\leq i,j\leq N.
\end{eqnarray}
Such coercive conditions restrict the selection of the trial function set 
and always generate extra computational work.

To overcome the first difficulty, the good performances of the artificial 
neural network (NN) solving high-dimensional partial differential equations (PDEs) 
had received a lot of attentions 
\cite{weinan2020machine,yu2018deep,han2017overcoming,raissi2019physics}.
These type of methods provides a possible way to solve many useful high-dimensional 
PDEs from physics, chemistry,
biology, engineers and so on \cite{daneker2022systems, han2019solving,raissi2019physics}.
Naturally, many fully-connected NN (FNN) based methods are applied to 
solve Schr\"{o}dinger equation by approximating
wave function via FNN architecture.
Among these applications, due to the universal approximation property \cite{hornik1989multilayer,hornik1990universal},
FNN can always provide a sufficient trial function set of wave function 
within the tolerable number of parameters. But neither the high-dimensional 
FNN itself nor the corresponding energy integration of FNN is easy
to implement direct quadrature scheme.
Therefore, Monte-Carlo method is always adopted to do these high-dimensional integration.
Monte-Carlo method is an inspiring idea to bypass the CoD and can be naturally 
combined with stochastic gradient descent method \cite{yu2018deep}, 
but in exchange for the extra uncertainty in the whole algorithm process.
For the bottleneck task such as solving the Schr\"{o}dinger equation, some necessary sampling
methods need to be considered \cite{han2019solving}.

To overcome the second difficulty, Slater determinant structure \cite{slater1930note} is the most widely
used way to ensure the anti-symmetry property.
There are many methodologies developed based on this structure such as Hartree-Fock (HF) 
based methods \cite{roothaan1960self,pople1954self}.
Recently, there are lots of attempts that combine the Slater determinant with the artificial neural 
network \cite{han2019solving,hermann2020deep,pfau2020ab}.
In these studies, the NN is constructed with a Slater determinant-like part which ensure the all elements
in trial function set satisfy the anti-symmetry property.
This extra part will generate additional number of nodes of the NN, and give rise to
computational complexity in the forward and backward propagation.

In this paper, the main idea that finding a way out of the two difficulties mentioned above is using 
a type of tensor neural network (TNN) to build the trial function set. Under TNN architecture, 
the high-precision direct quadrature rule, like tensor product Gauss quadrature rule, 
can be preformed in each terms of (\ref{Energy_Definition}) instead of using Monte-Carlo method.
In our previous work \cite{wang2022tensor}, we introduce the TNN architecture, 
prove the universal approximation property and show that the computational 
work for the integration of the functions built by TNN is only polynomial 
scale of the dimension. This means that TNN can be another idea to bypass 
the CoD and has the potential to work in solving high-dimensional 
PDE such as many-body Schr\"{o}dinger equation.
The most important property of TNN is that the corresponding high-dimensional 
functions can be easily integrated with high accuracy and high efficiency. 
Then, the deduced machine learning method can achieve high accuracy in solving high-dimensional problems. 
The reason is that the integration of TNN functions can be separated into one-dimensional 
integration which can be computed by classical quadrature schemes with high accuracy.   
The TNN  has been used to solve 20,000 dimensional Schr\"{o}dinger equation 
with coupled quantum harmonic oscillator potential function \cite{HuShuklaKarniadakisKawaguchi}, 
high-dimensional Fokker-Planck equations \cite{WangHuKawaguchiZhangKarniadakis} 
and high-dimensional time-dependent problems \cite{KaoZhaoZhang}. 
Since the high accuracy of high-dimensional integration of TNN functions, 
in this paper, the anti-symmetry property is ensured by adding penalty terms, 
which are the inner product of the wave functions before and after exchanging 
electron positions, to the loss function.
Since $\widehat H$ is spin-independent, for simplicity, we assume that the 
first $N_{\uparrow}$ electrons are of spin-up and the remaining 
$N_{\downarrow}=N-N_{\uparrow}$ electrons are of spin-down \cite{foulkes2001quantum}.
Since the aim here is to compute the ground state of the system,
we can assume $N_{\uparrow}=\lceil\frac{N}{2}\rceil$, where $\lceil k\rceil$ 
the smallest integer not less than $k$.
Then we modify the objective function as follows
\begin{eqnarray}\label{Energy_Penalty}
O[\Psi] = \frac{\langle \Psi|\widehat H|\Psi\rangle}{\langle \Psi|\Psi\rangle}
+\sum_{i=1}^{N_{\uparrow}}\sum_{j=i+1}^{N_{\uparrow}} \lambda_{ij}^{\uparrow}
\frac{\langle T_{ij}\Psi|\Psi\rangle }{ \langle\Psi|\Psi\rangle }
+\sum_{i=1}^{N_{\downarrow}}\sum_{j=i+1}^{N_{\downarrow}} \lambda_{ij}^{\downarrow}
\frac{\langle T_{ij}\Psi|\Psi\rangle }{ \langle\Psi|\Psi\rangle },
\end{eqnarray}
where $\lambda_{ij}^{\uparrow}$ and $\lambda_{ij}^{\downarrow}$ are the Lagrange 
multipliers corresponding to spin-up and spin-down electrons, respectively.

As preliminary tests, we show that our TNN-based method is able to 
solve the ground state electronic structure of atomic system without 
the pre-information of the reference energy and any information of atomic or molecular
orbital. Furthermore, the method here gives a potential to 
compute  the ground state for the molecules without any pre-information.
The method in this paper provides a possible way for solving large scale 
many-body Schr\"{o}dinger equations directly with high accuracy.

An outline of the paper goes as follows. In Section \ref{Section_TNN}, 
we introduce the TNN architecture. The numerical integration method 
for the functions built by TNN is designed in Section \ref{Section_Integration}.
Section \ref{Section_Eigenvalue} is devoted to proposing the TNN-based machine 
learning method for computing the ground state of many-body Schr\"{o}dinger equation. 
In Section \ref{Section_Efficient_Computation}, several types of efficient numerical methods 
are built for computing the loss function by designing TNN interpolations or discrete TNN 
expansion for Coulomb potentials. 
Some numerical examples are provided in Section \ref{Section_Numerical}
to show the validity and accuracy of the proposed numerical methods for computing the
electronic structure and molecular structure of diatomic molecules.
Some concluding remarks are given in the last section.

\section{Tensor neural network architecture}\label{Section_TNN}
TNN structure, its approximation property and the computational 
complexity of related integration have been detailedly discussed in \cite{wang2022tensor}.
In order to express clearly and facilitate the construction of the TNN method for 
solving Schr\"{o}dinger equation,
in this section, we will also elaborate on some important definitions and properties.

The TNN is constructed with $d$ subnetworks and each subnetwork
is a continuous mapping from a bounded closed set $\Omega_i\subset\mathbb R$ to $\mathbb R^p$,
which can be expressed as:
\begin{eqnarray}\label{def_FNN}
\phi_i(x_i;\vartheta_i)=\big(\phi_{i,1}(x_i;\vartheta_i),\phi_{i,2}(x_i;\vartheta_i),
\cdots,\phi_{i,p}(x_i;\vartheta_i)\big)^T,\ \ \ i=1, \cdots, d,
\end{eqnarray}
where each $x_i$ denotes the one-dimensional input, $\theta_i$ denotes 
the parameters of the $i$-th subnetwork, typically the weights and biases.
In this paper, the FNN architecture is chosen for building the  subnetworks.

After building all subnetworks, we combine the output layers of each subnetwork to
construct TNN architecture by the following mapping from $\R^d$ to $\R$
\begin{eqnarray}\label{def_TNN}
\Psi(x;\vartheta)=\sum_{j=1}^p\phi_{1,j}(x_1;\vartheta_1)
\phi_{2,j}(x_2;\vartheta_2)\cdots\phi_{d,j}(x_d;\vartheta_d)
=\sum_{j=1}^p\prod_{i=1}^d\phi_{i,j}(x_i;\vartheta_i),
\end{eqnarray}
where $x=(x_1,\cdots,x_d)\in\Omega_1\times\cdots\times\Omega_d$,
and $\vartheta=\{\vartheta_1,\cdots,\vartheta_d\}$ denotes the set of all trainable parameters.
In this paper, we simply assume $\Omega=\Omega_1\times\cdots\times\Omega_d$.
This setting of the calculation domain is reasonable for many high-dimensional physical problems.
\begin{figure}[htb]
\centering
\includegraphics[width=16cm,height=12cm]{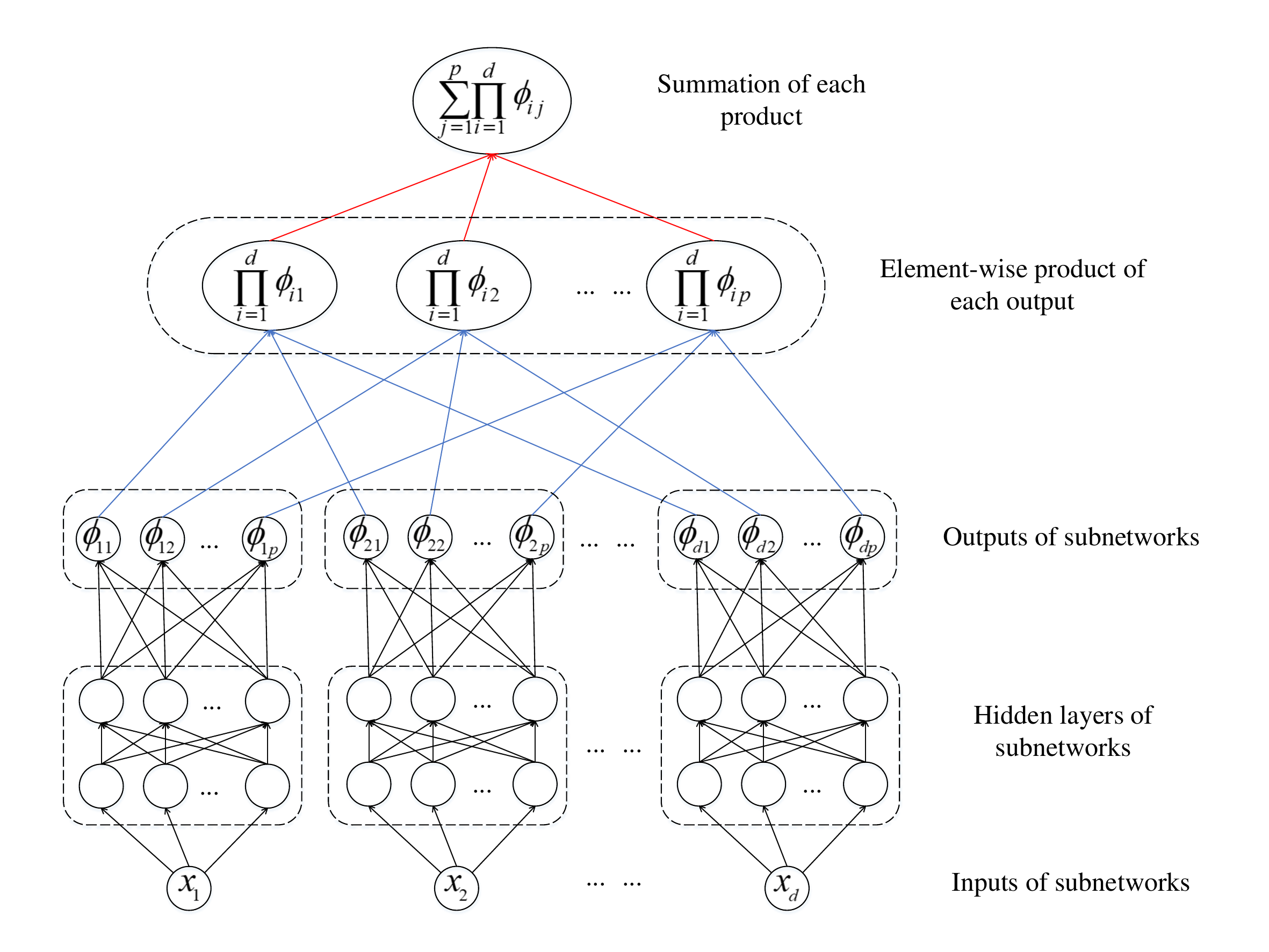}
\caption{Architecture of TNN. Black arrows mean linear transformation (or affine transformation). 
Each ending node of blue arrows is obtained by taking the scalar multiplication 
of all starting nodes of blue arrows that end in this ending node. 
The finall output of TNN is derived from summation of all starting nodes of red arrows.}
\end{figure}

In order to improve the numerical stability, we normalize each $\phi_{i,j}(x_i)$
and use the following normalized TNN structure:
\begin{eqnarray}\label{def_TNN_normed}
\Psi(x;c,\vartheta)&=&\sum_{j=1}^pc_j\widehat\phi_{1,j}(x_1;\vartheta_1)
\cdots\widehat\phi_{d,j}(x_d;\vartheta_d)=\sum_{j=1}^pc_j\prod_{i=1}^d\widehat\phi_{i,j}(x_i;\vartheta_i),
\end{eqnarray}
where each $c_j$ is a scaling parameter which describes 
the length of each rank-one function, $c=\{c_j\}_{j=1}^{p}$ 
is a set of trainable parameters, $\{c,\vartheta\}=\{c,\vartheta_1,\cdots,\vartheta_d\}$
denotes all parameters of the whole architecture.
For $i=1,\cdots,d,j=1,\cdots,p$, $\widehat\phi_{i,j}(x_i;\vartheta_i)$ is a 
$L^2$-normalized function as follows:
\begin{eqnarray*}
\widehat\phi_{i,j}(x_i;\vartheta_i)
=\frac{\phi_{i,j}(x_i;\vartheta_i)}{\|\phi_{i,j}(x_i;\vartheta_i)\|_{L^2(\Omega_i)}}.
\end{eqnarray*}
For simplicity of notation, $\phi_{i,j}(x_i;\vartheta_i)$ 
denotes the normalized function in the following parts.

Since the isomorphism relation between $L^2(\Omega_1\times\cdots\times\Omega_d)$
and the tensor product space $L^2(\Omega_1)\otimes\cdots\otimes L^2(\Omega_d)$ \cite{ryan2002introduction},
the process of approximating the function $f(x)\in L^2(\Omega_1\times\cdots\times\Omega_d)$
with the TNN defined by (\ref{def_TNN}) is actually to search 
an approximation $f(x)$ in the space $L^2(\Omega_1)\otimes\cdots\otimes L^2(\Omega_d)$ 
with the rank being not greater than $p$.
Due to the low-rank structure, we will find that the polynomial mapping acting on the 
TNN and its derivatives can be integrated on the tensor-product domain with small scale computational work.
In order to show the validity of solving PDEs by the TNN,  we introduce the
following approximation result to the functions of the space $L^2(\Omega_1\times\cdots\times\Omega_d)$
in the sense of $H^m(\Omega)$-norm. For more information, please check \cite{wang2022tensor}.
\begin{theorem}\cite{wang2022tensor}\label{theorem_approximation}
Assume that each $\Omega_i$ is a bounded closed interval in $\mathbb R$ for $i=1, \cdots, d$, 
$\Omega=\Omega_1\times\cdots\times\Omega_d$, and the function $f(x)\in H^m(\Omega)$. 
Then for any tolerance $\varepsilon>0$, 
there exist a positive integer $p$ and the corresponding TNN defined by (\ref{def_TNN}) 
such that the following approximation property holds
\begin{equation}\label{eq:L2_app}
\|f(x)-\Psi(x;\vartheta)\|_{H^m(\Omega)}<\varepsilon.
\end{equation}
\end{theorem}
For the Schr\"{o}dinger equation, it is enough to consider the case $m=1$ in this paper.

\section{Quadrature scheme for TNN}\label{Section_Integration}
For easily understanding the way we're dealing with the integration of 
the kinetic energy and the Coulomb potential energy in the Schr\"{o}dinger equation, 
in this section, we introduce the method to 
compute the numerical integration for polynomial composite function of TNN and its derivatives.
We will find that each integration in Section \ref{Section_Eigenvalue} can fits this scheme naturally.
The reader may refer to \cite{wang2022tensor} for more details.

We will show that the application of TNN can bring a significant reduction of
the computational complexity for the related numerical integration.
For the description, we introduce the following sets of multiple indices
\begin{eqnarray}
\mathcal B&:=&\Big\{\beta=(\beta_1,\cdots,\beta_d)\in\mathbb N_0^d\ \Big|\ |\beta|
:=\sum_{i=1}^d\beta_i\leq m \Big\},\\
\mathcal A&:=&\Big\{\alpha=(\alpha_\beta)_{\beta\in\mathcal B}\in\mathbb N_0^{|\mathcal B|}\
\Big|\ |\alpha|:=\sum_{\beta\in\mathcal B}\alpha_\beta\leq k \Big\},
\end{eqnarray}
where $\mathbb N_0$ denotes the set of all the non-negative integers,  $m$ and $k$ are two positive integers,
$|\mathcal B|$ and $|\mathcal A|$ denote the cardinal numbers of $\mathcal B$ and $\mathcal A$, respectively.
Here, we only focus on the high-dimensional cases where $m\ll d$ and $k\ll d$.
Simple calculation leads to the following equations
$$|\mathcal B|=\sum_{j=0}^m\binom{j+d-1}{j}, \ \ \ \ |\mathcal A|=\sum_{j=0}^k\binom{j+|\mathcal B|-1}{j}.$$
It is easy to know that the scales of magnitudes of $|\mathcal B|$ 
and $|\mathcal A|$ are $\mathcal O\big((d+m)^m\big)$ and $\mathcal O\big(((d+m)^m+k)^k\big)$, respectively.

In the following parts of this paper, the parameter $\vartheta$ in (\ref{def_TNN}) 
will be omitted for brevity without confusion. The activation function of TNN is chosen to be smooth enough
such that $\Psi(x)$ has partial derivatives up to order $m$.
Here, we assume $F(x)$ is built by the $k$-degree complete polynomial 
of $d$-dimensional TNN and its partial derivatives up to order $m$ that can be expressed as follows
\begin{eqnarray}\label{def_F(x)}
F(x)=\sum_{\alpha\in\mathcal A}A_{\alpha}(x)\prod_{\beta\in\mathcal B}
\left(\frac{\partial^{|\beta|}\Psi(x)}{\partial x_1^{\beta_1}
\cdots\partial x_d^{\beta_d}}\right)^{\alpha_\beta},
\end{eqnarray}
where the coefficient $A_\alpha(x)$ is defined by the following expansion
such that the rank of $A_\alpha(x)$ is not greater than $q$ in the tensor 
product space $L^2(\Omega_1)\otimes\cdots\otimes L^2(\Omega_d)$
\begin{eqnarray}\label{def_A_alpha}
A_\alpha(x)=\sum_{\ell=1}^q B_{1,\ell,\alpha}(x_1)B_{2,\ell,\alpha}(x_2)\cdots B_{d,\ell,\alpha}(x_d).
\end{eqnarray}
Here $B_{i,\ell,\alpha}(x_i)$ denotes the one-dimensional function in $L^2(\Omega_i)$ 
for $i=1, \cdots, d$ and  $\ell=1,\cdots, q$.
The essential idea to reduce the computational complexity of the numerical integration
$\int_\Omega F(x)dx$ is that the TNN function $F(x)$ can be decomposed into a tensor product structure.

In order to implement  the  decomposition, for each 
$\alpha = (\alpha_1, \cdots, \alpha_{|\mathcal B|})\in\mathcal A$, we give the following definition
\begin{eqnarray}
\mathcal B_\alpha:=\Big\{\beta=\left(\beta_1,\cdots,\beta_d\right)
\in\mathcal B\ \big|\ \alpha_\beta\geq 1\Big\}.
\end{eqnarray}
With the help of the index set $\mathcal A$, we can deduce that 
$|\mathcal B_\alpha|\leq k$ for any $\alpha\in\mathcal A$.

Since  $\Psi(x)$ has the TNN structure (\ref{def_TNN}), the cumprod can be further decomposed as
\begin{eqnarray}\label{eq_decomposition_prod}
&&\prod_{\beta\in\mathcal B_\alpha}
\left(\frac{\partial^{|\beta|}\Psi(x)}{\partial x_1^{\beta_1}
\cdots\partial x_d^{\beta_d}}\right)^{\alpha_\beta}
=\prod_{\beta\in\mathcal B_\alpha}\left(\frac{\partial^{|\beta|}\sum\limits_{j=1}^p
\phi_{1,j}(x_1)\cdots\phi_{d,j}(x_d)}{\partial x_1^{\beta_1}
\cdots\partial x_d^{\beta_d}}\right)^{\alpha_\beta}\nonumber\\
&&=\prod_{\beta\in\mathcal B_\alpha}\left(\sum\limits_{j=1}^p
\frac{\partial^{\beta_1}\phi_{1,j}(x_1)}{\partial x_1^{\beta_1}}
\cdots\frac{\partial^{\beta_d}\phi_{d,j}(x_d)}{\partial x_d^{\beta_d}} \right)^{\alpha_\beta}\nonumber\\
&&=\prod_{\beta\in\mathcal B_\alpha}\sum_{1\leq j_1,\cdots,j_{\alpha_\beta}\leq p}
\left(\frac{\partial^{\beta_1}\phi_{1,j_1}(x_1)}{\partial x_1^{\beta_1}}
\cdots\frac{\partial^{\beta_1}\phi_{1,j_{\alpha_{\beta}}}(x_1)}{\partial x_1^{\beta_1}}\right) \cdots\left(\frac{\partial^{\beta_d}\phi_{d,j_1}(x_d)}{\partial x_d^{\beta_d}}
\cdots\frac{\partial^{\beta_d}\phi_{d, j_{\alpha_{\beta}}}(x_d)}{\partial x_d^{\beta_d}}\right)\nonumber\\
&&=\prod_{\beta\in\mathcal B_\alpha}\sum_{1\leq j_1,\cdots,j_{\alpha_\beta}\leq p}\left(\prod_{\ell=1}^{\alpha_\beta}\frac{\partial^{\beta_1}\phi_{1,j_\ell}(x_1)}{\partial x_1^{\beta_1}}\right)
\cdots\left(\prod_{\ell=1}^{\alpha_\beta}
\frac{\partial^{\beta_d}\phi_{d,j_\ell}(x_d)}{\partial x_d^{\beta_d}}\right) \nonumber\\
&&=\sum_{\substack{\beta\in\mathcal B_\alpha, \ell=1,\cdots,\alpha_\beta, \\ 1\leq j_{\beta,\ell}\leq p}}
\left(\prod_{\beta\in\mathcal B_\alpha}\prod_{\ell=1}^{\alpha_\beta}
\frac{\partial^{\beta_1}\phi_{1,j_{\beta,\ell}}(x_1)}{\partial x_1^{\beta_1}}\right)
\cdots\left(\prod_{\beta\in\mathcal B_\alpha}\prod_{\ell=1}^{\alpha_\beta}
\frac{\partial^{\beta_d}\phi_{d,j_{\beta,\ell}}(x_d)}{\partial x_d^{\beta_d}}\right).
\end{eqnarray}
With the help of expansion (\ref{eq_decomposition_prod}),  
we can give the following expansion for $F(x)$
\begin{eqnarray}\label{Expansion_Fx}
F(x) &=& \sum_{\alpha\in\mathcal A}
\left(\sum_{\ell=1}^qB_{1,\ell,\alpha}(x_1)\cdots B_{d,\ell,\alpha}(x_d)\right)\nonumber\\
&&\ \ \ \cdot\sum_{\substack{\beta\in\mathcal B_\alpha, \ell=1,\cdots,\alpha_\beta,\\ 
1\leq j_{\beta,\ell}\leq p}}
\left(\prod_{\beta\in\mathcal B_\alpha}\prod_{\ell=1}^{\alpha_\beta}
\frac{\partial^{\beta_1}\phi_{1,j_{\beta,\ell}}(x_1)}{\partial x_1^{\beta_1}}\right)
\cdots\left(\prod_{\beta\in\mathcal B_\alpha}\prod_{\ell=1}^{\alpha_\beta}
\frac{\partial^{\beta_d}\phi_{d,j_{\beta,\ell}}(x_d)}{\partial x_d^{\beta_d}}\right)\nonumber\\
&=&\sum_{\alpha\in\mathcal A}\sum_{\ell=1}^q
\sum_{\substack{\beta\in\mathcal B_\alpha, \ell=1,\cdots,\alpha_\beta,\\ 1\leq j_{\beta,\ell}\leq p}}
\left(B_{1,\ell,\alpha}(x_1)\prod_{\beta\in\mathcal B_\alpha}\prod_{\ell=1}^{\alpha_\beta}
\frac{\partial^{\beta_1}\phi_{1,j_{\beta,\ell}}(x_1)}{\partial x_1^{\beta_1}}\right)\nonumber\\
&&\ \quad\quad \ \cdots\left(B_{d,\ell,\alpha}(x_d)
\prod_{\beta\in\mathcal B_\alpha}\prod_{\ell=1}^{\alpha_\beta}
\frac{\partial^{\beta_d}\phi_{d,j_{\beta,\ell}}(x_d)}{\partial x_d^{\beta_d}}\right).
\end{eqnarray}
Based on the decomposition (\ref{Expansion_Fx}), we have the following 
splitting scheme for the integration $\int_\Omega F(x)dx$
\begin{eqnarray}
\int_\Omega F(x)dx &=&\sum_{\alpha\in\mathcal A}\sum_{\ell=1}^q
\sum_{\substack{\beta\in\mathcal B_\alpha, \ell=1,\cdots,\alpha_\beta,\\ 1\leq j_{\beta,\ell}\leq p}}
\int_{\Omega_1}\left(B_{1,\ell,\alpha}(x_1)\prod_{\beta\in\mathcal B_\alpha}\prod_{\ell=1}^{\alpha_\beta}
\frac{\partial^{\beta_1}\phi_{1,j_{\beta,\ell}}(x_1)}{\partial x_1^{\beta_1}}\right)dx_1\nonumber\\
&&\ \quad\quad \ \cdots\int_{\Omega_d}\left(B_{d,\ell,\alpha}(x_d)
\prod_{\beta\in\mathcal B_\alpha}\prod_{\ell=1}^{\alpha_\beta}
\frac{\partial^{\beta_d}\phi_{d,j_{\beta,\ell}}(x_d)}{\partial x_d^{\beta_d}}\right)dx_n.
\end{eqnarray}

It is time to introduce the detailed numerical integration method for the TNN function $F(x)$.
Without loss of generality, for $i=1, \cdots, d$, we select $N_i$ Gauss points $\{x_i^{(n_i)}\}_{n_i=1}^{N_i}$
and the corresponding weights $\{w_i^{(n_i)}\}_{n_i=1}^{N_i}$ 
for the $i$-th dimensional domain $\Omega_i$,
and denote $\bar N=\max\{N_1,\cdots,N_d\}$.
Introducing the index $n=(n_1,\cdots,n_d)\in\mathcal N:=\{1,\cdots,N_1\}\times\cdots\times\{1,\cdots,N_d\}$,
the tensor product Gauss points and their corresponding weights on the domain $\Omega$ 
can be expressed as follows
\begin{eqnarray}\label{def_Gauss}
\left.
\begin{array}{rcl}
\Big\{x^{(n)}\Big\}_{n\in\mathcal N}&=&\left\{x_1^{(n_1)}\right\}_{n_1=1}^{N_1}
\times\left\{x_2^{(n_2)}\right\}_{n_2=1}^{N_2}
\times\cdots\times
\left\{x_d^{(n_d)}\right\}_{n_d=1}^{N_d},\\
\Big\{w^{(n)}\Big\}_{n\in\mathcal N}&=&\left\{\prod\limits_{i=1}^dw_i^{(n_i)}\ \Big| \ w_i^{(n_i)}\in\left\{w_i^{(n_i)}\right\}_{n_i=1}^{N_i},i=1,\cdots,d\right\}.
\end{array}
\right.
\end{eqnarray}
Fortunately, with the help of expansion (\ref{eq_decomposition_prod}),
we can give the following splitting numerical quadrature scheme for $\int_\Omega F(x)dx$:
\begin{eqnarray}\label{eq_I_tensor_form}
\int_\Omega F(x)dx&\approx&
\sum_{\alpha\in\mathcal A}\sum_{\ell=1}^q\sum_{\substack{\beta\in\mathcal B_\alpha, 
\ell=1,\cdots,\alpha_\beta,\\ 1\leq j_{\beta,\ell}\leq p}}
\left(\sum_{n_1=1}^{N_1}w_1^{(n_1)}B_{1,\ell,\alpha}(x_1^{(n_1)})
\prod_{\beta\in\mathcal B_\alpha}\prod_{\ell=1}^{\alpha_\beta}
\frac{\partial^{\beta_1}\phi_{1,j_{\beta,\ell}}(x_1^{(n_1)})}{\partial x_1^{\beta_1}}\right)\nonumber\\
&&\ \ \quad \ \cdots\left(\sum_{n_d=1}^{N_d}w_d^{(n_d)}B_{d,\ell,\alpha}(x_d^{(n_d)})
\prod_{\beta\in\mathcal B_\alpha}\prod_{\ell=1}^{\alpha_\beta}
\frac{\partial^{\beta_d}\phi_{d,j_{\beta,\ell}}(x_d^{(n_d)})}{\partial x_d^{\beta_d}}\right).
\end{eqnarray} 
The aim to design the quadrature scheme (\ref{eq_I_tensor_form}) is to decompose the
high-dimensional integration $\int_\Omega F(x)dx$ into to a series of one-dimensional integration.
The scheme  (\ref{eq_I_tensor_form}) can reduce the computational work of the 
high-dimensional integration for the $d$-dimensional function $F(x)$ to the polynomial 
scale of dimension $d$ due to the simplicity of the
one-dimensional integration. 
The following theorem shows the low computational complexity for the high dimension 
TNN functions. For more information, please refer to \cite{wang2022tensor}.
\begin{theorem}\cite{wang2022tensor}\label{theorem_Gauss}
Assume that the function $F(x)$ is defined by (\ref{def_F(x)}) with the coefficient 
$A_\alpha(x)$ having the expansion (\ref{def_A_alpha}).
On the $d$-dimensional tensor product domain $\Omega$, we choose the tensor product 
Gauss points and their corresponding weights which
are defined by (\ref{def_Gauss}) to determine the quadrature scheme.
Based on these Gauss points and weights, let us perform the numerical integration (\ref{eq_I_tensor_form})
for the function $F(x)$ on the domain $\Omega$. Let $T_1$ denote the computational complexity for the $1$-dimensional
function evaluation operations.

If the function $\Psi(x)$ involved in the  function $F(x)$ has the TNN form (\ref{def_TNN}),
the computational complexity for the numerical integration (\ref{eq_I_tensor_form}) can be bounded by
$\mathcal O\big(dqT_1k^2p^k\big((d+m)^m+k\big)^kN\big)$, which is the polynomial scale of the dimension $d$.
\end{theorem}

\section{Solving Schr\"{o}dinger equation by TNN}\label{Section_Eigenvalue}
In this section, we introduce the application of TNN for computing the ground state of the
many-body Schr\"{o}dinger equation (\ref{Energy_Definition}) by using
the TNN-based machine learning method.

For the description of the numerical method, let us state the the energy definition for the
Hamiltonian (\ref{Hamiltonian_Operator_2})
\begin{eqnarray}\label{Energy}
\langle \Psi|\widehat H|\Psi\rangle = \frac{1}{2}\sum_{i=1}^N\int |\nabla_i\Psi|^2d\mathbf r 
+ \sum_{i=1}^N\sum_{j=i+1}^N \int \frac{|\Psi|^2}{|\mathbf r_i-\mathbf r_j|}d\mathbf r
- \sum_{I=1}^M\sum_{i=1}^N\int \frac{Z_I|\Psi|^2}{|\mathbf r_i-\mathbf R_I|}d\mathbf r.
\end{eqnarray}
In order to compute the singular integrals of the Coulomb potential terms in (\ref{Energy}),
the spherical coordinates $(r, \theta, \varphi)$ are adopted here.
Then the wave function $\Psi(\mathbf r)$ should be written as 
$\Psi(r_1,\theta_1, \varphi_1$, $\cdots$, $r_N,\theta_N,\varphi_N)$.
The Laplace $\Delta$ has following expression
\begin{eqnarray}\label{laplace}
\Delta\Psi&=&\frac{\partial^2\Psi}{\partial r^2}+\frac{2}{r}\frac{\partial\Psi}{\partial r}
+\frac{1}{r^2}\frac{\partial^2\Psi}{\partial\theta^2}
+\frac{\cos\theta}{r^2\sin\theta}\frac{\partial\Psi}{\partial\theta}
+\frac{1}{r^2\sin^2\theta}\frac{\partial^2\Psi}{\partial\varphi^2}\nonumber\\
&=&\frac{1}{r^2}\frac{\partial}{\partial r}\left(r^2\frac{\partial\Psi}{\partial r}\right)
+\frac{1}{r^2\sin\theta}\frac{\partial}{\partial\theta}
\left(\sin\theta\frac{\partial\Psi}{\partial\theta}\right)
+\frac{1}{r^2\sin^2\theta}\frac{\partial^2\Psi}{\partial\varphi^2}.
\end{eqnarray}
In order to do the integration for the term 
$\int \frac{|\Psi|^2}{|\mathbf r_i-\mathbf r_j|}d\mathbf r$ in (\ref{Energy}),
we introduce the expansion for $1/|\mathbf r_i-\mathbf r_j|$ 
by the associated Legendre polynomials \cite{arfken1985mathematical}.
For this aim, it is enough to consider the expansion for the term $1/r_{12}:=1/|\mathbf r_1-\mathbf r_2|$.
Let us define the $\ell$-th order Legendre polynomial $P_\ell (x)$ in the interval $x\in[-1, 1]$.
When $\ell$ is a positive integer, the $\ell$-th order Legendre polynomial is defined 
by the Rodrigues formula
\begin{eqnarray}\label{Equality_1}
P_\ell (x)=\frac{1}{2^\ell \ell!}\frac{\partial^\ell}{\partial x^\ell}(x^2-1)^\ell,
\end{eqnarray}
which can be computed by the following recursive formula
\begin{eqnarray}\label{Equality_2}
P_\ell (x)=\frac{2\ell-1}{\ell}xP_{\ell-1}(x)-\frac{\ell-1}{\ell}P_{\ell-2}(x).
\end{eqnarray}
When $\ell$ is a nonnegative integer and  $|m|\leq \ell$,  the associated Legendre polynomials $P_\ell ^m(x)$
can be generated from the Legendre polynomial $P_\ell (x)$ by the following way
\begin{eqnarray}\label{Legendre_to_ALegendre}
P_\ell ^m(x)&=&\left(1-x^2\right)^{\frac{m}{2}}\frac{{\rm d}^m}{{\rm d}x^m}P_\ell (x)
=\frac{(1-x^2)^{\frac{m}{2}}}{2^\ell \ell!}
\frac{{\rm d}^{\ell+m}}{{\rm d}x^{\ell+m}}\left(x^2-1\right)^\ell.
\end{eqnarray}
Finally, based on the definitions of (\ref{Equality_1}), (\ref{Equality_2}) and (\ref{Legendre_to_ALegendre}),
 the term $1/|\mathbf r_1-\mathbf r_2|$ can be expanded as follows
\begin{eqnarray}\label{min and max}
\frac{1}{r_{12}}&=&\sum_{\ell=0}^\infty\frac{r_<^\ell}{r_>^{\ell+1}}\Big[P_\ell (\cos\theta_1)
P_\ell (\cos\theta_2)\nonumber\\
&&+2\sum_{m=1}^\ell\frac{(\ell-m)!}{(\ell+m)!}P_\ell^m(\cos\theta_1)
P_\ell^m(\cos\theta_2)\cos(m(\varphi_1-\varphi_2) )\Big]\nonumber\\
&=&\sum_{\ell=0}^\infty\frac{r_<^\ell}{r_>^{\ell+1}}\Big[P_\ell (\cos\theta_1)
P_\ell (\cos\theta_2)\nonumber\\
&&+2\sum_{m=1}^\ell\frac{(\ell-m)!}{(\ell+m)!}P_\ell^m(\cos\theta_1)
P_\ell^m(\cos\theta_2)\cos (m\varphi_1)\cos (m\varphi_2)\nonumber\\
&&+2\sum_{m=1}^\ell\frac{(\ell-m)!}{(\ell+m)!}P_\ell^m(\cos\theta_1)
P_\ell^m(\cos\theta_2)\sin (m\varphi_1)\sin( m\varphi_2) \Big],
\end{eqnarray}
where $\mathbf r_1=(r_1,\theta_1, \varphi_1)$, $\mathbf r_2=(r_2,\theta_2,\varphi_2)$,  
$r_>=\max\{r_1, r_2\}$ and $r_<=\min\{r_1, r_2\}$.
The important property of the expansion (\ref{min and max}) is that the variables $r_1$, $\theta_1$,
$\varphi_1$, $r_2$, $\theta_2$, $\varphi_2$  in the expansion are
separable except the functions $r_>$ and $r_<$ which will be discussed in Section \ref{Section_Efficient_Computation}.
In practical numerical computation, we will truncate the expansion of 
$1/|\mathbf r_1-\mathbf r_2|$ into finite $n$ terms as follows
\begin{eqnarray}\label{Expansion_n}
\frac{1}{r_{12}^{(n)}}&=&\sum_{\ell=0}^n\frac{r_<^\ell}{r_>^{\ell+1}}
\Big[P_\ell (\cos\theta_1)P_\ell (\cos\theta_2)\nonumber\\
&&+2\sum_{m=1}^\ell\frac{(\ell-m)!}{(\ell+m)!}P_\ell^m(\cos\theta_1)
P_\ell^m(\cos\theta_2)\cos (m\varphi_1)\cos (m\varphi_2)\nonumber\\
&&+2\sum_{m=1}^\ell\frac{(\ell-m)!}{(\ell+m)!}P_\ell^m(\cos\theta_1)
P_\ell^m(\cos\theta_2)\sin (m\varphi_1)\sin( m\varphi_2) \Big].
\end{eqnarray}
The terms $1/r_{ij}:=1/|\mathbf r_i-\mathbf r_j|$ and $r_{Ii} := 1/|\mathbf r_i-\mathbf R_I|$
can be done the same truncations as (\ref{Expansion_n}) for $1/r_{12}$.
Based on these truncations, in practical computation, we use the following truncated Hamiltonian operator
\begin{eqnarray}\label{Hamilton_n}
\widehat H_n = -\frac{1}{2}\sum_{i=1}^N \nabla_i^2  
+\sum_{i=1}^N\sum_{j=i+1}^N\frac{1}{r_{ij}^{(n)}} 
-\sum_{I=1}^M\sum_{i=1}^N\frac{Z_I}{r_{Ii}^{(n)}},
\end{eqnarray}
where $1/r_{ij}^{(n)}$ and $1/r_{Ii}^{(n)}$ denote the truncations of 
the terms $1/r_{ij}$ and $1/r_{Ii}$, respectively.

In the following parts of this paper, for the simple description of
the TNN for problem (\ref{Energy_Definition}),
we use $x$ to denote the coordinates 
$(r_1,\theta_1, \varphi_1,\cdots, r_N,\theta_N,\varphi_N)$ and $x_i$ 
denotes one coordinate component. 
In order to solve the optimization problem (\ref{Energy_Definition}),
we build a TNN structure $\Psi(x;\vartheta)$ by the way (\ref{def_TNN}),
and define the set of all possible values of $\vartheta$ as $\Theta$.
For $i=1,\cdots, d$, the subnetworks $\phi_i(x_i;\vartheta_i),i=1,\cdots,d$ 
are defined as (\ref{def_FNN}). We choose sufficient smooth 
activation functions, such that $\Psi(x;\vartheta)\in H_0^1(\Omega)$ with the
open boundary conditions.

The trial wave function  set $V$ is modeled by the TNN structure $\Psi(x;\vartheta)$ 
where parameters $\vartheta$ take all the possible values 
and it is obvious that $V\subset H_0^1(\Omega)$. The solution $\Psi(x;\vartheta^*)$ 
of the following optimization problem is the approximation to the ground state wave function :
\begin{eqnarray}\label{approx_opt}
\Psi(x,\vartheta^*)&=&\arg\min_{\Psi(x;\vartheta)\in V}
\frac{\langle \Psi(x;\vartheta)| \widehat H_n|\Psi(x;\vartheta)\rangle}{\langle \Psi(x;\vartheta)|
\Psi(x;\vartheta)\rangle}+\sum_{i=1}^{N_{\uparrow}}
\sum_{j=i+1}^{N_{\uparrow}} \lambda_{ij}^{\uparrow}\frac{\langle T_{ij}\Psi(x;\vartheta)|
\Psi(x;\vartheta)\rangle}
{\langle\Psi(x;\vartheta)|\Psi(x;\vartheta)\rangle }\nonumber\\
&&+\sum_{i=1}^{N_{\downarrow}}\sum_{j=i+1}^{N_{\downarrow}} \lambda_{ij}^{\downarrow}
\frac{\langle T_{ij}\Psi(x;\vartheta)|\Psi(x;\vartheta)\rangle }{ \langle\Psi(x;\vartheta)|
\Psi(x;\vartheta)\rangle }.
\end{eqnarray}

Note that all integrals of the numerator and the denominator of (\ref{approx_opt}) 
have the form (\ref{def_F(x)}). With the help of Theorem \ref{theorem_Gauss},
 these numerical integration can be implemented by the scheme (\ref{eq_I_tensor_form})
 with the computational work being bounded by the polynomial scale of dimension $d=3N$.
We choose the tensor product Gauss points and their corresponding weights which are defined
by (\ref{def_Gauss}) to discretize these numerical integration.
Then  the loss function can be defined as follows
\begin{eqnarray}\label{loss}
L(\vartheta)&:=&\frac{\sum\limits_{n\in\mathcal N}w^{(n)}|\Psi(x^{(n)},\vartheta) 
\widehat H_n \Psi(x^{(n)},\vartheta) |
}{\sum\limits_{n\in\mathcal N}\Psi^2(x^{(n)};\vartheta)} 
+ \sum_{i=0}^{N_{\uparrow}}\sum_{j=i+1}^{N_{\uparrow}}\lambda_{ij}^{\uparrow}
\frac{\sum\limits_{n\in \mathcal N}\big(T_{ij}\Psi(x^{(n)};\vartheta)\big)
\Psi(x^{(n)};\vartheta)}{\sum\limits_{n\in\mathcal N}\Psi^2(x^{(n)};\vartheta)}\nonumber\\
&& + \sum_{i=0}^{N_{\downarrow}}\sum_{j=i+1}^{N_{\downarrow}}\lambda_{ij}^{\downarrow}
\frac{\sum\limits_{n\in \mathcal N}\big(T_{ij}\Psi(x^{(n)};\vartheta)\big)
\Psi(x^{(n)};\vartheta)}{\sum\limits_{n\in\mathcal N}
\Psi^2(x^{(n)};\vartheta)},
\end{eqnarray}
where all integrals are computed by the quadrature scheme (\ref{eq_I_tensor_form}).

In this paper, the gradient descent (GD) method is adopted to minimize 
the loss function $L(\vartheta)$. The GD scheme can be described as follows:
\begin{eqnarray}\label{gd}
\vartheta^{(k+1)}=\vartheta^{(k)}-\eta\nabla L(\vartheta^{(k)}),
\end{eqnarray}
where $\vartheta^{(k)}$ denotes the parameters after the $k$-th GD  step, 
$\eta$ is the learning rate (step size).
In practical learning process, we use Adam optimizer \cite{kingma2014adam} with 
adaptive learning rates and L-BFGS to get the optimal solution $\Psi(x;\vartheta^*)$.

\section{Efficient computation for the loss function}\label{Section_Efficient_Computation}
This section is devoted to designing efficient algorithm for computing the energy \eqref{loss}, 
especially for the integration involved the $n$-term expansion \eqref{Expansion_n}. 
Since the term $\frac{r_{<}^{\ell}}{r_{>}^{\ell+1}}$ in \eqref{Expansion_n} is basically 
non-separable, special treatment must be applied to this non-separable function 
so as to further improve the computing efficiency for the loss function \eqref{loss}. 
In this section, we introduce two types of way to achieve this, 
namely to build the variable-separated approximating version for \eqref{Expansion_n}, 
and to explore the structure of two-dimensional discrete values 
to accelerate the involved two-dimensional integration.

\subsection{TNN interpolation}\label{Section_Interpolation}
Since the two dimensional function \( \frac{r_{<}^{\ell}}{r_{>}^{\ell+1}} \) 
in (\ref{Expansion_n}) represents the primary 
singularity and exhibits poor separability, it cannot be directly expressed in the separable format 
of TNN. Compared with the one dimensional integration, the complexity of two-dimensional Gaussian 
integration is quite high and computationally expensive. Based on the universal approximation 
capability of TNNs \cite{wang2022tensor}, it is natural idea to employ TNN interpolation \cite{LiLinWangXie} 
techniques to approximate \( \frac{r_{<}^{\ell}}{r_{>}^{\ell+1}} \) to achieve a variable separable approximation, 
which can then be substituted into the expression $\frac{r_{<}^{\ell}}{r_{>}^{\ell+1}}r_1^2r_2^2$, 
transforming it into a product of one-dimensional Gaussian integrals in separable form
\begin{eqnarray}\label{TNN_Interpolation}
\Phi^{(\ell)}(r_1,r_2)  =  \sum_{j=1}^{p_{r}^{(\ell)}}\alpha_{j}^{(\ell)}
\prod_{i=1}^{2} \psi_{i,j}^{(\ell)}(r_i)\longrightarrow\frac{r_{<}^{\ell}}{r_{>}^{\ell+1}}r_1^2r_2^2. 
\end{eqnarray}

The reason for employing TNN interpolation for $\frac{r_{<}^{\ell}}{r_{>}^{\ell+1}}r_1^2r_2^2$ is that incorporating 
the \( \frac{r_{<}^{\ell}}{r_{>}^{\ell+1}} \) with the  term \( r_1^2 r_2^2 \) helps control the overall singularity, 
since it exhibits a uniform bound with respect to the parameter \( \ell \). 
This owns a significant advantage during the approximation process, 
as it localizes the singularity to the line \( y = x \). This characteristic facilitates 
the subsequent use of adaptive interpolation points to address the singularity. 
However, a drawback is the extension of the function's range. 
For example, for the wave function of the ground state helium atom, 
the radial truncation interval in polar coordinates is \([0, 5]\), 
but the range of this function sometimes need to be extended to \([0, 125]\). 
This extension leads to a rapid increase of the derivatives for the objective function as \( \ell \) increases. 
To address this issue, we come to introduce a coordinate transformation. 
As an example, for the helium atom, the Schr\"{o}dinger equation in spherical coordinates is given by: 
\begin{equation}
-\frac{1}{2}\Delta\Psi - \frac{2\Psi}{r_1} - \frac{2\Psi}{r_2} + \frac{\Psi}{r_{12}} = E\Psi,
\end{equation}
where \(\Psi = \Psi(r_1, \theta_1, \varphi_1, r_2, \theta_2, \varphi_2)\). 
Here, the distance term  $r_{12}$ is defined as $r_{12} = |\mathbf r_1-\mathbf r_2|$, 
where $\mathbf r_1 = (r_1,\theta_1,\varphi_1)$, $\mathbf r_2 = (r_2,\theta_2,\varphi_2)$. 
By using the coordinate transformation as \(r_{i}=s\cdot  t_i\),  
we can define the tensor neural network to approximate the wave function as follows
\begin{eqnarray}
&&\Psi(s \cdot t_1, \theta_1, \varphi_1, s \cdot t_2, \theta_2, \varphi_2)\nonumber\\
&&= \sum_{j=1}^p \alpha_{j}\phi_{t_1,j}(s \cdot t_1) \phi_{t_2,j}(s \cdot t_2) 
\cdot  \phi_{\theta_1,j}(\theta_1) \phi_{\theta_2,j}(\theta_2) \cdot  \phi_{\varphi_1,j}(\varphi_1) 
\phi_{\varphi_2,j}(\varphi_2).
\end{eqnarray}
We are interested in finding the ground state of the helium atom, which satisfies 
the Courant-Fischer min-max theorem. 
In numerical computations, we truncate the integration domain 
to a bounded region $D  \subset \mathbb{R}^6$. 
The region \(D\) in spherical coordinates can be transformed to the domain  
\(\widetilde{D} = [0,1]^2\times [0,\pi]^2\times[0,2\pi]^2\). 
Then the ground state energy $E_1$ can be computed as follows  
\begin{eqnarray}\label{Energy_1}
&&E_1 = \inf_{\Psi} \frac{ \frac{1}{2}\int_{D} |\nabla \Psi|^2dx 
- 2\int_{D} \frac{\Psi^2}{r_1} \, dx -2 \int_{D} \frac{\Psi^2}{r_2} \, dx 
+ \int_{D} \frac{\Psi^2}{r_{12}} \, dx}{\int_{D} \Psi^2 \, dx}\nonumber\\
&&=\inf_{\Psi(t;\theta)} \frac{\frac{1}{2}\int_{\widetilde{D}} |\nabla \Psi(t;\theta)|^2dt 
- 2s\int_{\widetilde{D}} \frac{\Psi^2(t;\theta)}{t_1}dt 
- 2s\int_{\widetilde{D}} \frac{\Psi^2(t;\theta)}{t_2}dt 
+ s\int_{\widetilde{D}} \frac{\Psi^2(t;\theta)}{t_{12}}dt}{s^2\int_{\widetilde{D}} \Psi^2(t;\theta)dt},
\end{eqnarray}
where the constant $s={\rm radial}(D)/{\rm radial}(\widetilde D)$.   

By employing the variable separation property of TNN, the integration $\int_{D}\Psi^{2}dD$ in (\ref{Energy_1}) 
can be efficiently computed by using the one-dimensional Gaussian quadrature scheme:
\begin{eqnarray}
\int_{D}\Psi^{2}dD&=& s^6 \sum_{i = 1}^{p}\sum_{j = 1}^{p}\alpha_i\alpha_j\int_0^1t_{1}^{2}
\phi_{t_{1},i}(s\cdot  t_{1})\phi_{t_{1},j}(s\cdot  t_{1})dt_{1} \cdot \int_0^1t_{2}^{2}
\phi_{t_{2},i}(s\cdot  t_{2})\phi_{t_{2},j}(s\cdot t_{2})dt_{2}\nonumber\\
&&\cdot \int_0^{\pi}\sin\theta_{1}\phi_{\theta_{1},i}(\theta_{1})
\phi_{\theta_{1},j}(\theta_{1})d\theta_{1}\cdot  \int_0^{\pi}\sin\theta_{2}\phi_{\theta_{2},i}(\theta_{2})
\phi_{\theta_{2},j}(\theta_{2})d\theta_{2}\nonumber \\
&&\cdot \int_0^{2\pi}\phi_{\varphi_{1},i}(\varphi_{1})\phi_{\varphi_{1},j}(\varphi_{1})d\varphi_{1}
\cdot \int_0^{2\pi}\phi_{\varphi_{2},i}(\varphi_{2})\phi_{\varphi_{2},j}(\varphi_{2})d\varphi_{2}.
\end{eqnarray}
Based on the Laplace operator's representation  (\ref{laplace}) in polar coordinates, 
the kinetic energy term can be transformed via the coordinate change into the following form with separated variables:
\begin{eqnarray}
\int_{D}|\nabla\Psi|^2dD &=&s^4 \int_{\widetilde{D}}t_{1}^{2}t_{2}^{2}\sin\theta_{1}\sin\theta_{2}
\left[\Big(\frac{\partial\Psi}{\partial t_{1}}\Big)^{2}+\frac{1}{t_{1}^{2}}
\Big(\frac{\partial\Psi}{\partial\theta_{1}}\Big)^{2}
+\frac{1}{t_{1}^{2}\sin^{2}\theta_{1}}\Big(\frac{\partial\Psi}{\partial\varphi_{1}}\Big)^{2}\right.\nonumber\\
&&+\left.\Big(\frac{\partial\Psi}{\partial t_{2}}\Big)^{2}+\frac{1}{t_{2}^{2}}
\Big(\frac{\partial\Psi}{\partial\theta_{2}}\Big)^{2}
+\frac{1}{t_{2}^{2}\sin^{2}\theta_{2}}\Big(\frac{\partial\Psi}{\partial\varphi_{2}}\Big)^{2}\right]d\widetilde D.
\end{eqnarray}
Similarly, applying the scale transformation to the coordinates, the Coulomb potential term 
$\int_{D}\frac{\Psi^{2}}{r_1}dD$ can be immediately expressed as the following separated variable form: 
\begin{eqnarray}
\int_{D}\frac{\Psi^{2}}{r_1}dD &=&s^5 \sum_{i=1}^{p}\sum_{j=1}^{p}\alpha_i\alpha_j\int_0^1t_{1}\phi_{t_{1},i}(t_{1})
\phi_{t_{1},j}(t_{1})dt_{1}\cdot \int_0^1t_{2}^{2}\phi_{t_{2},i}(t_{2})\phi_{t_{2},j}(t_{2})dt_{2}\nonumber\\
&&\cdot \int_0^{\pi}\sin\theta_{1}\phi_{\theta_{1},i}(\theta_{1})\phi_{\theta_{1},j}(\theta_{1})d\theta_{1}
\cdot \int_0^{\pi}\sin\theta_{2}\phi_{\theta_{2},i}(\theta_{2})\phi_{\theta_{2},j}(\theta_{2})d\theta_{2}\nonumber\\
&&\cdot \int_0^{2\pi}\phi_{\varphi_{1},i}(\varphi_{1})\phi_{\varphi_{1},j}(\varphi_{1})d\varphi_{1}\cdot 
\int_0^{2\pi}\phi_{\varphi_{2},i}(\varphi_{2})\phi_{\varphi_{2},j}(\varphi_{2})d\varphi_{2}.
\end{eqnarray}
Then the term $\int_{D}\frac{\Psi^{2}}{r_2}dD$ can be computed similarly. 

Combining the spherical harmonics and associated Legendre polynomials (\ref{Expansion_n}), 
TNN approximation (\ref{TNN_Interpolation}) to $\frac{r_{<}^{\ell}}{r_{>}^{\ell+1}}r_1^2r_2^2$, 
the following integration formulas hold:
\begin{eqnarray}\label{truncation}
&&\int_{D}\frac{\Psi^{2}}{r_{12}}\,dD = s^5 \sum_{i = 1}^{p}
\sum_{j = 1}^{p}\alpha_i\alpha_j\sum_{\ell=0}^{n}\sum_{k=1}^{p_r^{(\ell)}}
\alpha_k^{(\ell)}\int_0^1\psi_{t_1,k}^{(\ell)}(t_1)\phi_{t_{1},i}(s t_{1})\phi_{t_{1},j}(s t_{1})\,dt_{1}\nonumber \\
&&\cdot \int_0^1\psi_{t_2,k}^{(\ell)}(t_2)\phi_{t_{2},i}(s t_{2})\phi_{t_{2},j}(s t_{2})\,dt_{2} \nonumber\\
&&\cdot  \left[ \int_0^{\pi}\sin\theta_{1}P_\ell(\cos\theta_{1})\phi_{\theta_{1},i}(\theta_{1})
\phi_{\theta_{1},j}(\theta_{1})d\theta_{1}
\cdot  \int_0^{\pi}\sin\theta_{2}P_\ell(\cos\theta_{2})
\phi_{\theta_{2},i}(\theta_{2})\phi_{\theta_{2},j}(\theta_{2})d\theta_{2} \right.\nonumber\\
&&\cdot  \left. \int_0^{2\pi}\phi_{\varphi_{1},i}(\varphi_{1})\phi_{\varphi_{1},j}(\varphi_{1})\,d\varphi_{1} 
\cdot  \int_0^{2\pi}\phi_{\varphi_{2},i}(\varphi_{2})\phi_{\varphi_{2},j}(\varphi_{2})\,d\varphi_{2} \right.\nonumber\\
&&\left. + 2\sum_{m=1}^{\ell}\frac{(\ell-m)!}{(\ell+m)!}\int_0^{\pi}\sin\theta_{1}P_\ell^m(\cos\theta_{1})
\phi_{\theta_{1},i}(\theta_{1})\phi_{\theta_{1},j}(\theta_{1})d\theta_{1}
\right.\nonumber\\
&&\cdot \int_0^{\pi}\sin\theta_{2}P_\ell^m(\cos\theta_{2})\phi_{\theta_{2},i}(\theta_{2})
\phi_{\theta_{2},j}(\theta_{2})d\theta_{2} \underline{}\cdot
\left.\int_0^{2\pi}\cos(m\varphi_{1})\phi_{\varphi_{1},i}(\varphi_{1})
\phi_{\varphi_{1},j}(\varphi_{1})\,d\varphi_{1}\right.\nonumber\\
&&\cdot  \int_0^{2\pi}\cos(m\varphi_{2})\phi_{\varphi_{2},i}(\varphi_{2})
\phi_{\varphi_{2},j}(\varphi_{2})d\varphi_{2}\nonumber \\
&& + 2\sum_{m=1}^{\ell}\frac{(\ell-m)!}{(\ell+m)!}\int_0^{\pi}\sin\theta_{1}P_\ell^m(\cos\theta_{1})
\phi_{\theta_{1},i}(\theta_{1})\phi_{\theta_{1},j}(\theta_{1})\,d\theta_{1}\nonumber\\
&&\cdot  \int_0^{\pi}\sin\theta_{2}P_\ell^m(\cos\theta_{2})\phi_{\theta_{2},i}(\theta_{2})
\phi_{\theta_{2},j}(\theta_{2})\,d\theta_{2} \cdot  \left. \int_0^{2\pi}\sin(m\varphi_{1})
\phi_{\varphi_{1},i}(\varphi_{1})\phi_{\varphi_{1},j}(\varphi_{1})\,d\varphi_{1} \right.\nonumber\\
&&\cdot \left. \int_0^{2\pi}\sin(m\varphi_{2})\phi_{\varphi_{2},i}(\varphi_{2})
\phi_{\varphi_{2},j}(\varphi_{2})\,d\varphi_{2} \right].
\end{eqnarray}
By using the coordinate transformation, we can use TNN interpolation techniques to approximate 
$\frac{r_{<}^{\ell}}{r_{>}^{\ell+1}}r_{1}^{2}r_{2}^{2}$ in the region $[0,1]^{2}$ 
due to its universal approximation ability \cite{wang2022tensor}. 

We use the same TNN interpolation approximation method as in \cite{LiLinWangXie}. 
For the detailed approximation process, at each iteration $m$, 
we obtain a bunch of training points $x_{k}^{(m)} := (x_{k,1}^{(m)}, \cdots, x_{k,d}^{(m)})^{\top}$, 
$k=1, \cdots, K$ according to sampling rules, and minimize the squared loss
\begin{eqnarray}
L_{m}(\Theta)&:=&\sum_{k=1}^{K}\left(\Psi(x_{k}^{(m)},\Theta)-f(x_{k}^{(m)})\right)^{2}\nonumber\\
&=&\sum_{k=1}^K\left(\sum_{j=1}^pc_j\prod_{i=1}^d\widehat{\phi}_{i,j}(x_{k,i}^{(m)};\vartheta_i)
-f(x_k^{(m)})\right)^2, 
\end{eqnarray}
to obtain the desired network parameters $\Theta = \{c, \vartheta_1, \cdots, \vartheta_d \}$. 
This procedure is repeated $M$ times until obtain good enough results on the validation scheme, 
such as the accuracy on the set of test points. 

In each training process, we split the parameters into two groups $\{c\}$ and $\{\vartheta_1, \cdots, \vartheta_d\}$. 
The parameter $c$ can be regarded as the linear coefficients on the $p$-dimensional subspace 
$$V_{p}^{(m)} := {\rm span}\left\{\psi_{j}(x;\vartheta^{(m)}) 
:= \prod_{i = 1}^{d} \widehat{\phi}_{i,j}(x_i ; \vartheta_i^{(m)})\right\}.$$
And therefore, we only need to solve a linear equation to obtain the optimal coefficient $c$ 
on the current subspace $V_p^{(m)}$ in the sense of the squared loss on training points. 
Using the optimal coefficient $c$, we update the network parameters $\{\vartheta_j\}$ by minimizing 
the loss function with some optimization algorithm. The detailed TNN interpolation method 
to obtain an approximation for a given target function is defined in Algorithm \ref{alg}.
\begin{algorithm}[ht!]
\caption{TNN interpolation method}\label{alg}
\SetAlgoNlRelativeSize{0}
\SetAlgoNlRelativeSize{-1}
\SetAlgoNlRelativeSize{-2}
\SetAlgoNlRelativeSize{1}
\SetAlgoNlRelativeSize{2}
\SetAlgoNlRelativeSize{-1}
    
\KwIn{Target function $f(x)$, TNN function $\Psi(x;\Theta)$ defined by (3.1), 
initial model parameters $\Theta$, domain $\Omega$.}
\KwOut{Learned approximate TNN function $\Psi ( x; \Theta ^* )$.}
\KwData{Number of total training iterations $M$, number of training points in each iteration $K$, 
number of optimization steps $T$ for each training points set, hyper-parameters of optimization 
algorithm such as learning rate $\gamma$.}

\For{$m \leftarrow 1 $ \textbf{to} $M$}{
    Sample $x_{k}^{( m ) }\in \Omega$, $k= 1, \cdots, K$ according to some sampling rules.
                
    \For{$t\leftarrow 1$ \textbf{to} $T$}{
        Assemble matrix $A^{( t) }$ and vector $B^{( t) }$  as follows:
        \[
        \begin{aligned}
        &A_{\mu,\nu}^{(t)}=\sum_{k=1}^{K}\prod_{i=1}^{d}
        \widehat{\phi}_{i,\mu}\left(x_{k,i}^{(m)};\vartheta_{i}^{(t-1)}\right)
        \prod_{i=1}^{d}\widehat{\phi}_{i,\nu}\left(x_{k,i}^{(m)};\vartheta_{i}^{(t-1)}\right),\ \ \:1\leq \mu,\nu\leq p,\\
        &B_{\mu}^{(t)}=\sum_{k=1}^{K}f(x_{k}^{(m)})\prod_{i=1}^{d}\widehat{\phi}_{i,\mu}\left(x_{k,i}^{(m)};
        \vartheta_{i}^{(t-1)}\right),\ \:1\leq \mu\leq p.
        \end{aligned}
        \]
        Solve the following linear equation to obtain the solution $c\in\mathbb{R}^p$:
        \[
        A^{(t)}c=B^{(t)},
        \]
        and update the coefficient parameter as $c^{(t)}=c.$
        
        Compute the loss function:
        \[
        \mathcal{L}_m^{(t)}(\vartheta^{(t-1)})=\sum_{k=1}^K\left(\sum_{j=1}^pc_j^{(t)}
        \prod_{i=1}^d\widehat{\phi}_{i,j}\left(x_{k,i}^{(m)};\vartheta_i^{(t-1)}\right)-f(x_k^{(m)})\right)^2.
        \]
        
        Use an optimization step to update the neural network parameters of TNN as follows:
        \[
        \vartheta^{(t)}=\vartheta^{(t-1)}
        -\gamma\frac{\partial\mathcal{L}_m^{(t)}}{\partial\vartheta}(c^{(t)},\vartheta^{(t-1)}).
        \]
    }
}
\end{algorithm} 
It is noteworthy that the model derived from directly interpolating 
the function $\frac{r_{<}^{\ell}}{r_{>}^{\ell+1}}r_{1}^{2}r_{2}^{2}$ is highly efficient in practice. 
This approach not only requires minimal memory but also allows for eigenvalue calculations 
on a personal computer once the interpolation is completed. 
Additionally, this method has a broad applicability: all singular terms involving Coulomb 
potentials can be represented in a similar form 
using associated Legendre polynomial expansions. The computational effort is predominantly concentrated 
on the TNN interpolation for the function $\frac{r_{<}^{\ell}}{r_{>}^{\ell+1}}r_{1}^{2}r_{2}^{2}$. 
However, once the interpolated data is obtained, it is possible to directly load it from the model, 
transforming the task of finding the ground state energy into TNN structures and a product of one-dimensional integrals. 

\subsection{Discrete tensor-product expansion}\label{Subsection_TNN_Gauss}
This subsection is dedicated to designing two types of methods to obtain the 
the discrete tensor expansion for \eqref{Expansion_n} at two-dimensional 
quadrature points, so that the involved two-dimensional integration 
can be efficiently computed. In fact, 
to enhance the efficiency for computing the loss function \eqref{loss}, 
it is necessary to transform the expansion \eqref{Expansion_n} into a variable-separated form.  

One of the key advantages of using TNNs to solve PDEs is their 
exceptional ability to separate variables, which enables high-dimensional integrals 
to be computed as products of several one-dimensional integrals. 


When computing the loss function for the Schr\"{o}dinger equation, 
the only term that remains non-separable in integration after 
expanding $\frac{1}{r_{12}}$ with spherical harmonic functions is a two-dimensional function 
\(\frac{r_{<}^{\ell}}{r_{>}^{\ell+1}}\). 

In order to find the essential difficulty for computing Coulomb potential, as an example,
let us also consider the calculation of the ground state energy of a helium atom, 
which involves the following term in the integral
\begin{eqnarray}\label{Potential_Energy}
&&\sum_{i=1}^{p}\sum_{j=1}^{p}\sum_{\ell=0}^{n}\int_{\mathbb R^2}r_{1}^{2}r_{2}^{2}
\frac{r_{<}^{\ell}}{r_{>}^{\ell+1}}
\phi_{r_{1},i}(r_{1})\phi_{r_{1},j}(r_{1})\phi_{r_{2},i}(r_{2})
\phi_{r_{2},j}(r_{2})dr_{1}dr_{2}\nonumber\\
&&\cdot \int_0^{\pi}\sin(\theta_1)P_\ell(\cos\theta_1)\phi_{\theta_1,i}(\theta_1)
\phi_{\theta_1,j}(\theta_1)d\theta_1\cdot 
\int_0^{\pi}\sin(\theta_2)P_\ell(\cos\theta_2)\phi_{\theta_2,i}(\theta_2)
\phi_{\theta_2,j}(\theta_2)d\theta_2\quad\quad\nonumber\\
&&\cdot \int_0^{2\pi}\phi_{\varphi_1,i}(\varphi_1)\phi_{\varphi_1,j}(\varphi_1)d\varphi_1\cdot 
\int_0^{2\pi}\phi_{\varphi_2,i}(\varphi_2)\phi_{\varphi_2,j}(\varphi_2)d\varphi_2,
\end{eqnarray}
where \(r_< =\min\{r_1,r_2\}\), \(r_>=\max\{r_1,r_2\}\), \(\ell\) is a constant, 
and the function \(r_{1}^{2}r_{2}^{2}\frac{r_{<}^{\ell}}{r_{>}^{\ell+1}}\) is not variable-separated. 

The main difficulty for computing (\ref{Potential_Energy}) is the two-dimensional integration over the 
$r_1\times r_2$ direction. For practical computation, we truncate the whole space $\mathbb R^2$ 
to the finite Cartesian-product-type region  \([0 , r_{1,\rm max}] \times [0 , r_{2,\rm max}]\). 
Now, we should come to consider the following two-dimensional integration
$$
\int_0^{r_{1,\rm max}}\int_0^{r_{2,\rm max}}r_{1}^{2}r_{2}^{2}\frac{r_{<}^{\ell}}{r_{>}^{\ell+1}}
\phi_{r_{1},i}(r_{1})\phi_{r_{1},j}(r_{1})\phi_{r_{2},i}(r_{2})\phi_{r_{2},j}(r_{2})dr_{1}dr_{2}.
$$

\subsubsection{Method 1: leveraging the separable structure on each subdomain}
In this subsection, we dig into the separable structure of $\frac{r_{<}^{\ell}}{r_{>}^{\ell+1}}$ on each subdomain.
Notice that the function \(r_{1}^{2}r_{2}^{2}\frac{r_{<}^{\ell}}{r_{>}^{\ell+1}}\) can be written  
as \(r_1^{1-\ell} r_2^{2+\ell}\) in the triangular region where \(r_1 \geq r_2\), 
and as \(r_2^{1-\ell} r_1^{2+\ell}\) in the region where \(r_1 \leq r_2\). 
And therefore, the function \(r_{1}^{2}r_{2}^{2}\frac{r_{<}^{\ell}}{r_{>}^{\ell+1}}\) 
can be represented in a variable-separated TNN structure over both triangular regions.

For simplicity and understanding, we set $r_{1,\max}=r_{2,\max}=:r_{\max}$ 
and choose the same quadrature scheme for $r_1$ and $r_2$. 
Let \(\mathcal{C} = \{p_m\}_{m=1}^{\bar N}\) be the set of one-dimensional Gaussian quadrature points 
and $\{\omega_m\}_{m=1}^{\bar N}$ the corresponding quadrature weights on the interval $[0, r_{\max}]$, 
where \(p_m \leq p_{m+1}\) for \(1 \leq m \leq \bar N-1\). Consider the TNN interpolating function 
with a rank parameter \(p = \bar N\), defined as follows
\begin{eqnarray}\label{Points_r1r2}
\Psi(r_1, r_2) &=& \Psi_{\rm upper}(r_1, r_2) + \Psi_{\rm lower}(r_1, r_2) \nonumber\\
&=& \sum_{m=0}^{\frac{\bar N}{2}-1} \psi_{r_1,m}^{\rm upper}(r_1) \cdot \psi_{r_2,m}^{\rm upper}(r_2) 
+ \sum_{k=0}^{\frac{\bar N}{2}-1} \psi_{r_1,k}^{\rm lower}(r_1) \cdot \psi_{r_2,k}^{\rm lower}(r_2),
\end{eqnarray}
where
\begin{eqnarray}\label{Function_Upper}
\Psi_{\rm upper}(r_1, r_2) &=& \sum_{m=0}^{\frac{\bar N}{2}-1} \psi_{r_1,m}^{\rm upper}(r_1) 
\cdot \psi_{r_2,m}^{\rm upper}(r_2)= I_{[0,p_2]}(r_1) r_1^{2+\ell} \cdot I_{[p_2,r_{\max}]}(r_2) r_2^{1-\ell}\nonumber \\
&&+ \sum_{m=1}^{\frac{\bar N}{2}-1} I_{(p_{2m},p_{2m+2}]}(r_1) r_1^{2+\ell} 
\cdot I_{[p_{2m+2},r_{\max}]}(r_2) r_2^{1-\ell},
\end{eqnarray}
and
\begin{eqnarray}\label{Function_Lower}
\Psi_{\rm lower}(r_1, r_2) &=& \sum_{k=0}^{\frac{\bar N}{2}-1} \psi_{r_1,k}^{\rm lower}(r_1) 
\cdot \psi_{r_2,k}^{\rm lower}(r_2)= I_{[p_1,r_{\max}]}(r_1) r_1^{1-\ell} \cdot I_{[0,p_1]}(r_2) r_2^{2+\ell} \nonumber\\
&&+ \sum_{k=1}^{\frac{\bar N}{2}-1} I_{(p_{2k+1},r_{\max}]}(r_1) r_1^{1-\ell} 
\cdot I_{(p_{2k-1},p_{2k+1}]}(r_2) r_2^{2+\ell}.
\end{eqnarray}
Here, \(I_\Omega(r_k)\), for \(k= 1, 2\), denotes the characteristic function over the region \(\Omega\) 
in the \(k\)-th dimension.

For all two-dimensional quadrature points \((r_1, r_2) \in \mathcal{C} \times \mathcal{C}\), 
according to the expression of the TNN interpolation function (\ref{Points_r1r2}), we have
\[
\Psi(r_1, r_2) = r_1^{2} r_2^{2} \frac{r_{<}^{\ell}}{r_{>}^{\ell+1}},
\]
at the two-dimensional quadrature points.
Then, we can compute the two dimensional integration as follows 
\begin{eqnarray}\label{Integration_2}
&& \int_0^{r_{\max}}\int_0^{r_{\max}} r_1^2 r_2^2 \frac{r_{<}^{\ell}}{r_{>}^{\ell+1}} 
\phi_{r_1, i}(r_1) \phi_{r_1, j}(r_1) 
\phi_{r_2, i}(r_2) \phi_{r_2, j}(r_2) \, dr_1 \, dr_2 \nonumber\\
&& \approx \sum_{\mu=1}^{\bar N} \sum_{\nu=1}^{\bar N} \omega_{\mu}\omega_{\nu} 
(p_{\mu})^2 (p_{\nu})^2 \frac{\min\{p_{\mu}, p_{\nu}\}^\ell}
{\max\{p_{\mu}, p_{\nu}\}^{\ell+1}} \phi_{r_1, i}(p_{\mu}) 
\phi_{r_1, j}(p_{\mu})\phi_{r_2, i}(p_{\nu}) \phi_{r_2, j}(p_{\nu}) \nonumber\\
&& = \sum_{\mu=1}^{\bar N} \sum_{\nu=1}^{\bar N} \omega_{\mu}\omega_{\nu} 
\cdot \Psi(p_{\mu}, p_{\nu}) \phi_{r_1, i}(p_{\mu}) \phi_{r_1, j}(p_{\mu})
\phi_{r_2, i}(p_{\nu})\phi_{r_2, j}(p_{\nu})\nonumber\\
&& \approx \int_0^{r_{\max}}\int_0^{r_{\max}} \Psi(r_1, r_2) 
\phi_{r_1, i}(r_1) \phi_{r_1, j}(r_1)\phi_{r_2, i}(r_2) \phi_{r_2, j}(r_2) \, dr_1 \, dr_2 \nonumber\\
&& = \sum_{m=0}^{\frac{\bar N}{2}-1} \int_0^{r_{\max}}\int_0^{r_{\max}}
\psi_{r_1,m}^{\rm upper}(r_1)\cdot \psi_{r_2,m}^{\rm upper}(r_2) \phi_{r_1, i}(r_1) 
\phi_{r_1, j}(r_1) \phi_{r_2, i}(r_2) \phi_{r_2, j}(r_2) \, dr_1 \, dr_2 \nonumber\\
&& + \sum_{k=0}^{\frac{\bar N}{2}-1} \int_0^{r_{\max}}\int_0^{r_{\max}} \psi_{r_1,k}^{\rm lower}(r_1) 
\cdot \psi_{r_2,k}^{\rm lower}(r_2) \phi_{r_1, i}(r_1) \phi_{r_1, j}(r_1) 
\phi_{r_2, i}(r_2)\phi_{r_2, j}(r_2)\, dr_1 \, dr_2 \nonumber\\
&& =\sum_{m=0}^{\frac{\bar N}{2}-1} 
\int_0^{r_{\max}} \psi_{r_1,m}^{\rm upper}(r_1) \phi_{r_1, i}(r_1) \phi_{r_1, j}(r_1) \, dr_1 
\int_0^{r_{\max}} \psi_{r_2,m}^{\rm upper}(r_2) \phi_{r_2, i}(r_2) \phi_{r_2, j}(r_2) \, dr_2 \nonumber\\
&& + \sum_{k=0}^{\frac{\bar N}{2}-1} \int_0^{r_{\max}} \psi_{r_1,k}^{\rm lower}(r_1) \phi_{r_1, i}(r_1) 
\phi_{r_1, j}(r_1) \, dr_1 \int_0^{r_{\max}} \psi_{r_2,k}^{\rm lower}(r_2) 
\phi_{r_2, i}(r_2) \phi_{r_2, j}(r_2) \, dr_2.\quad\quad\quad 
\end{eqnarray} 
By directly considering the numerical integration on the two dimensional discrete quadrature nodes, 
we have successfully achieved discrete variable separation for two-dimensional integrals with high accuracy. 

From the definition (\ref{Function_Upper}) and the property of characteristic function, it is easy to know that 
the following quadrature scheme holds for $m=0,\cdots, \frac{\bar N}{2}$
\begin{eqnarray}\label{Quadrature_Upper}
\int_0^{r_{\max}} \psi_{r_1,m}^{\rm upper}(r_1) \phi_{r_1, i}(r_1) \phi_{r_1, j}(r_1) \, dr_1
=\sum_{\mu=2m+1}^{2m+2} (p_\mu)^{2+\ell}\phi_{r_1,i}(p_\mu)\phi_{r_1,j}(p_\mu)\omega_{\mu},
\end{eqnarray}
which is only summation of two terms.  Similarly, from (\ref{Function_Lower}), we also have
\begin{eqnarray}
&&\int_0^{r_{\max}}\psi_{r_2,0}^{\rm lower}(r_2) 
\phi_{r_2, i}(r_2) \phi_{r_2, j}(r_2) \, dr_2 = (p_1)^{2+\ell}\phi_{r_2,i}(p_1)\phi_{r_2,j}(p_1)\omega_{1},\label{Quadrature_Lower_1}\\
&&\int_0^{r_{\max}}\psi_{r_2,k}^{\rm lower}(r_2) 
\phi_{r_2, i}(r_2) \phi_{r_2, j}(r_2) \, dr_2\nonumber\\
&&\quad\quad\quad = \sum_{\nu=2k}^{2k+1} (p_\nu)^{2+\ell}\phi_{r_2,i}(p_\nu)\phi_{r_2,j}(p_\nu)\omega_{\nu}, \ \ \ 
 k=1, \cdots, \frac{\bar N}{2}.\label{Quadrature_Lower}
\end{eqnarray}
From (\ref{Quadrature_Upper})-(\ref{Quadrature_Lower}), half of the 
one dimensional integrals in (\ref{Integration_2}) can be simplified into a sum of two terms.  
Then, the integration (\ref{Integration_2}) can be further simplified.

\subsubsection{Method 2: dealing with the two-dimensional data itself}
This subsection is dedicated to dealing with the two-dimensional 
data itself by expressing it in a different form. This new perspective 
allows us to design the optimal tensor summation order for computing 
the two-dimensional integration involved in Coulomb potentials of the loss function \eqref{loss}, 
and speeding up the whole training process. 

For the calculation of the Hamiltonian \eqref{Energy}, 
the term $1/r_{12}:=1/|\mathbf r_1-\mathbf r_2|$ 
is the obstacle to our efficient computation since it is highly non-separable. 
And that is the initial motivation for us to seek a separable expansion of this term 
for efficiently computing the involved energy. 
Using the associated Legendre expansion of $1/r_{12}:=1/|\mathbf r_1-\mathbf r_2|$ 
is an almost perfect way for our purpose. 
However, the term \( \frac{r_{<}^{\ell}}{r_{>}^{\ell+1}} \) in (\ref{Expansion_n}) 
remains non-separable and the two-dimensional integration seems to be inevitable, 
which becomes the speed bottleneck for computing the loss function. 
In Subsections \ref{Section_Interpolation} and \ref{Subsection_TNN_Gauss}, 
we develop two ways to approximate the non-separable term \( \frac{r_{<}^{\ell}}{r_{>}^{\ell+1}} \) 
using separable TNN functions and all that remains is one-dimensional integration. 
In this subsection, we deal with the two-dimensional integration directly 
using a trick adapted for PyTorch to accelerate computation. 
In our numerical test, we observe a rather satisfactory speed-up for computing the two-dimensional integration.

To illustrate our trick to deal with the two-dimensional integration, as an example,
let us also consider computing the integral (\ref{Potential_Energy}). 
The key point is to perform the following two-dimensional integration over a Cartesian 
product-type region \([0 , r_{1, \rm max}] \times [0 , r_{2, \rm max}]\)
$$
\int_0^{r_{1,\rm max}}\int_0^{r_{2,\rm max}}r_{1}^{2}r_{2}^{2}
\frac{r_{<}^{\ell}}{r_{>}^{\ell+1}}\phi_{r_{1},i}(r_{1})
\phi_{r_{1},j}(r_{1})\phi_{r_{2},i}(r_{2})\phi_{r_{2},j}(r_{2})dr_{1}dr_{2}.
$$
Let \(\mathcal{C}_1 := \{p_{1,m}\}_{m=1}^{N_1}\) 
and \(\mathcal{C}_2 := \{p_{2,k}\}_{k=1}^{N_2}\) 
be the set of one-dimensional Gaussian quadrature points on $r_1 \in [0 , r_{1,\rm max}]$ 
and $r_2 \in [0 , r_{2,\rm max}]$, respectively, 
\(\{w_{1,m}\}_{m=1}^{N_1}\) and \(\{w_{2,k}\}_{k=1}^{N_2}\) the corresponding quadrature weights. 
We conduct the two-dimensional Gauss-Legendre quadrature to compute the integration. 
Let us define $g(r_1, r_2):= r_{1}^{2}r_{2}^{2}\frac{r_{<}^{\ell}}{r_{>}^{\ell+1}}$ 
and compute the two-dimensional function values on the quadrature 
points \(\{(p_{1,m}, p_{2,k})\}_{m=1,\cdots, N_1, k=1, \cdots, N_2}\). 
The pointwise values can be expressed in the following form
\begin{equation}\label{discrete_TP_expansion}
g(r_1, r_2) = \sum_{m=1}^{N_1} I_{[r_1 = p_{1,m}]} g(p_{1,m}, r_2),  
\quad \text{for} \, r_1 \in \mathcal{C}_1, r_2 \in \mathcal{C}_2,
\end{equation}
where $I_{[r_1 = p_{1,m}]}$ is the indicator function of the single-point set $\{p_{1,m}\}$.
Then we can compute the two-dimensional integration as follows 
\[
\begin{aligned}
&\int_0^{r_{1,\rm max}}\int_0^{r_{2,\rm max}}r_{1}^{2}r_{2}^{2}
\frac{r_{<}^{\ell}}{r_{>}^{\ell+1}}
\phi_{r_{1},i}(r_{1})\phi_{r_{1},j}(r_{1})\phi_{r_{2},i}(r_{2})\phi_{r_{2},j}(r_{2})dr_{1}dr_{2}\\
= & \int_0^{r_{1,\rm max}}\int_0^{r_{2,\rm max}}g(r_1, r_2)
\phi_{r_{1},i}(r_{1})\phi_{r_{1},j}(r_{1})\phi_{r_{2},i}(r_{2})\phi_{r_{2},j}(r_{2})dr_{1}dr_{2}\\
= & \int_0^{r_{1,\rm max}}\int_0^{r_{2,\rm max}} 
\sum_{m=1}^{N_1} I_{[r_1 = p_{1,m}]} g(p_{1,m}, r_2) 
\phi_{r_{1},i}(r_{1})\phi_{r_{1},j}(r_{1})\phi_{r_{2},i}(r_{2})
\phi_{r_{2},j}(r_{2})dr_{1}dr_{2}\\
= & \sum_{m=1}^{N_1} \int_0^{r_{1,\rm max}} I_{[r_1 = p_{1,m}]}
\phi_{r_{1},i}(r_{1})\phi_{r_{1},j}(r_{1}) dr_1 \int_0^{r_{2,\rm max}}g(p_{1,m}, r_2)
\phi_{r_{2},i}(r_{2})\phi_{r_{2},j}(r_{2})dr_{2}\\
\approx & \sum_{m=1}^{N_1} \sum_{\mu=1}^{N_1} w_{1,\mu}I_{[p_{1,\mu} = p_{1,m}]}
\phi_{r_{1},i}(p_{1,\mu})\phi_{r_{1},j}(p_{1,\mu})
\sum_{k=1}^{N_2} w_{2,k} g(p_{1,m}, p_{2,k})\phi_{r_{2},i}(p_{2,k})\phi_{r_{2},j}(p_{2,k})\\
= &\sum_{m=1}^{N_1} w_{1,m} \phi_{r_{1},i}(p_{1,m})\phi_{r_{1},j}(p_{1,m}) 
\sum_{k=1}^{N_2} w_{2,k} g(p_{1,m}, p_{2,k})\phi_{r_{2},i}(p_{2,k})\phi_{r_{2},j}(p_{2,k}).
\end{aligned}
\]
Note that in \eqref{discrete_TP_expansion}, we express the multi-dimensional function 
values on the Cartesian product of one-dimensional points in a tensor-product form, 
similar to the structure of TNN function. This is achieved using 
indicator functions on single-point sets and piling up all the discrete 
values onto one dimension (in this case $r_2$). Instead of generating the 
function values of 
$\phi_{r_{1},i}(r_{1})\phi_{r_{1},j}(r_{1})\phi_{r_{2},i}(r_{2})\phi_{r_{2},j}(r_{2})$ 
over $\mathcal{C}_1 \times \mathcal{C}_2$ and performing the two-dimensional 
integration, we utilize the tensor-product structure of the TNN function, 
and eliminate one dimension of the sum at a time. 

It is easy to know that $g(r_1, r_2)$ on $\mathcal{C}_1 \times \mathcal{C}_2$ 
can also be expressed as 
\[
  g(r_1, r_2) = \sum_{k=1}^{N_2} g(r_1, p_{2,k}) I_{[r_2 = p_{2,k}]}, 
\]
where all the discrete values are `piled up' onto the $r_1$ dimension. 
This results in the computation order where the $r_1$ 
dimension is eliminated first in the sum. While eliminating 
one dimension at a time might seem natural and straightforward, 
expressing the discrete multi-dimensional point values in the form 
of \eqref{discrete_TP_expansion} allows for flexibility in designing 
different computation orders, enabling us to choose the most efficient 
one for our purpose. 
For example, suppose we are given the values of a four-dimensional function $h(r_1, r_2, r_3, r_4)$ 
on the discrete point set $\{p_{1,m}\}_{m=1}^{N_1} \times \{p_{2,k}\}_{k=1}^{N_2} 
\times \{p_{3,\mu}\}_{\mu=1}^{N_3} \times \{p_{4,\nu}\}_{\nu=1}^{N_4}$. 
If we want to compute a four-dimensional integral involving 
$h(r_1, r_2, r_3, r_4)$ using these discrete values, we can express
\[
h(r_1, r_2, r_3, r_4) = \sum_{m=1}^{N_1} \sum_{\nu=1}^{N_4} I_{[r_1=p_{1,m}]}
h(p_{1,m}, r_2, r_3, p_{4,\nu}) I_{[r_4=p_{4,\nu}]}.
\]
In this approach, during the actual computation, the two-dimensional integration 
over $r_2$ and $r_3$ is performed first, followed by the sequential elimination 
of the $r_1$ and $r_4$ dimensions. The optimal order of computation for 
multi-dimensional integration depends on factors such as memory layout, 
parallel computing strategies, and other considerations. In our numerical tests, 
the method outlined in \eqref{discrete_TP_expansion} has been highly effective, 
particularly in enabling efficient two-dimensional integration. 
In future work, we intend to apply this strategy by using TNN to solve the Kohn-Sham 
equations.


\section{Numerical experiments}\label{Section_Numerical}
In this section, we provide several examples to validate the efficiency and accuracy of the 
TNN-based machine learning method for solving Schr\"{o}dinger equation. 
First, we test the numerical performance of TNN interpolation for the kernel function $1/r_{12}$ 
of Coulomb potential. Then,  in the following subsections, the discrete TNN expansion 
is adopted to build the TNN based machine learning method for solving Schr\"{o}dinger equations 
of helium and lithium atoms, hydrogen molecule. 
\subsection{Numerical experiments for interpolation}
\begin{figure}[htbp]
\centering
\includegraphics[width=0.59\textwidth]{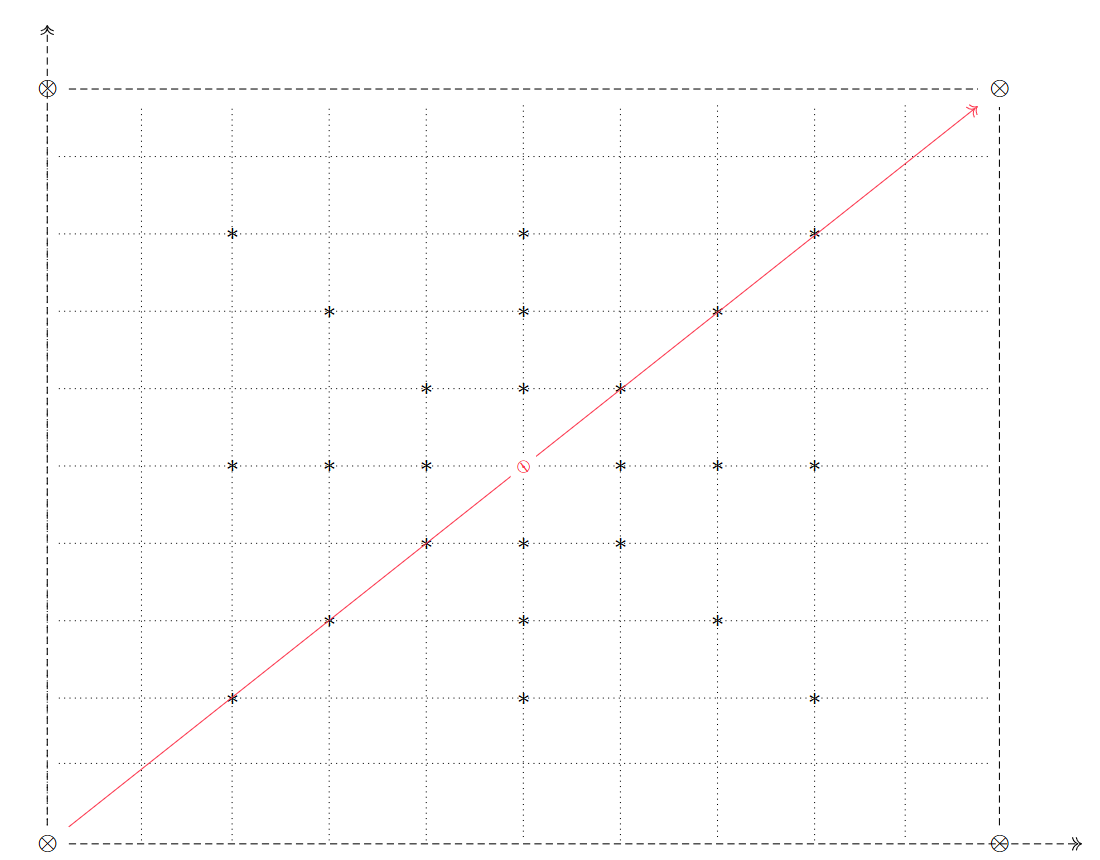}
\caption{Adaptively adding interpolation points}
\label{fig:adapt}
\end{figure}

In the first subsection, we investigate the performance of the TNN interpolation \cite{LiLinWangXie} 
techniques to approximate \(\frac{r_{<}^{\ell}}{r_{>}^{\ell+1}}r_1^2r_2^2\). 
Using Algorithm \ref{alg}, we propose the adaptive interpolation method to address the singularity at $y=x$. 
This method increases the density of interpolation points near the Gaussian points on the diagonal. 
Concretely, we add eight nodes surrounding each Gaussian point on the diagonal 
in each adaptive step, as illustrated in Figure \ref{fig:adapt}. 

For the radial direction of the wave function, the interval \([0,1]\) is uniformly divided into $25$ 
subintervals, 
with each subdivided interval containing $4$ quadrature points for the Gaussian 
quadrature. The selected interpolation points for training include both the aforementioned Gaussian 
nodes and additional nodes introduced by the adaptive strategy. 
However, test of interpolation errors uses only the Gaussian nodes. The training process involves 
alternating between the Adam and LBFGS steps. Specifically, 30,000 steps are performed 
using the Adam optimizer with a learning rate of \(9 \times 10^{-3}\), followed by 2,000 steps using the 
LBFGS optimizer with a learning rate of $1$. The results are summarized in Table~\ref{Table_1}.

\begin{table}[htb!]
\begin{center}
\caption{Interpolation Error for \( \frac{r_{<}^{\ell}}{r_{>}^{\ell+1}}r_1^2r_2^2\)}\label{Table_1}
\begin{tabular}{ccccc}
\hline
$\ell$& ${\rm Error}_{\max}$ & ${\rm Error}_{\rm abs-average}$ &\\
\hline
0&   0.000585367069217746&     2.4671756475080075e-05   \\
1&   0.001681433020818357&     6.8502735081223843e-05   \\
2&   0.002995919036041683&     0.00011106611019363546   \\
3&   0.003820942582003439&     0.00014025911765069184   \\
4&   0.004323711752593218&     0.00019276806158503203   \\
5&   0.005621666556553995&     0.00022419848555393435   \\
6&   0.005967029102881383&     0.00027846794433434256   \\
7&   0.007773735374113855&     0.00031117257239226967   \\
8&   0.007595491789484543&     0.00034901699692397800   \\
9&   0.008349611108816768&     0.00033828793430037285   \\
\hline
\end{tabular}
\end{center}
\end{table}

\begin{figure}[htb!]
\centering
\begin{minipage}{0.45\textwidth}
\centering
\includegraphics[width=\textwidth]{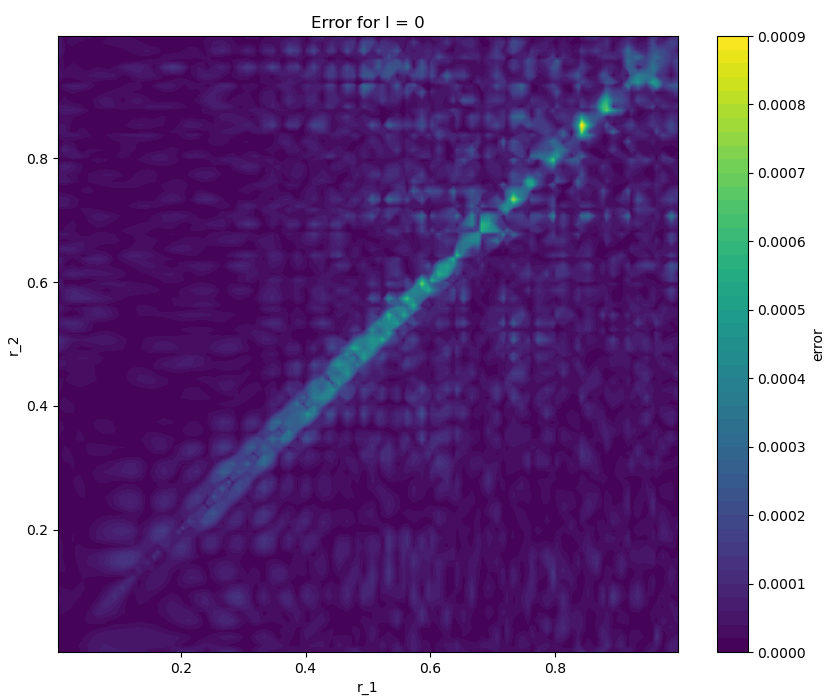}\label{fig:image1}
\end{minipage}\hspace{0.1pt}
\begin{minipage}{0.45\textwidth}
\centering
\includegraphics[width=\textwidth]{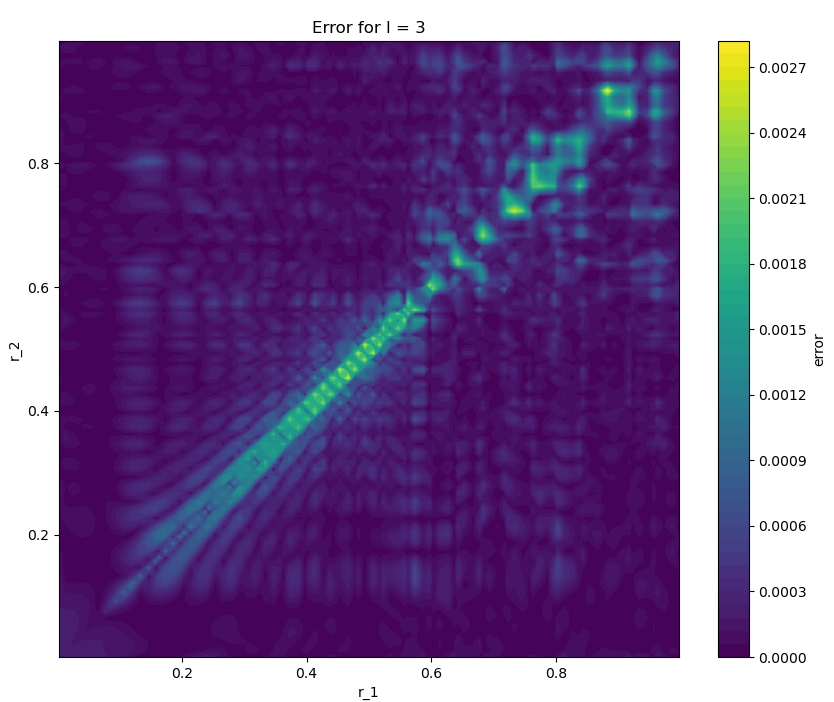}\label{fig:image2}
\end{minipage}
\\
\hspace{0.1pt} 
\begin{minipage}{0.45\textwidth}
\centering
\includegraphics[width=\textwidth]{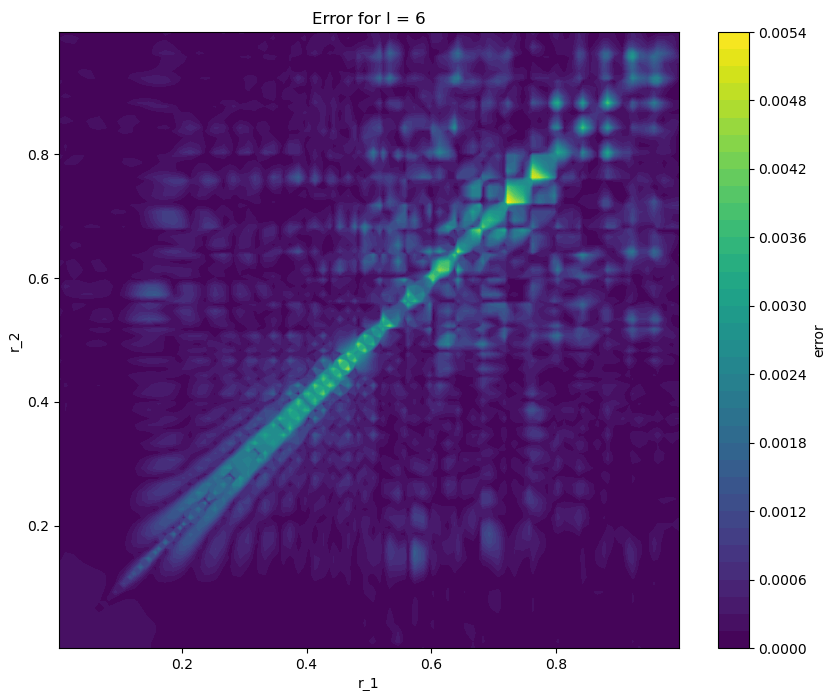}\label{fig:image3}
\end{minipage}\hspace{0.1pt}
\begin{minipage}{0.45\textwidth}
\centering
\includegraphics[width=\textwidth]{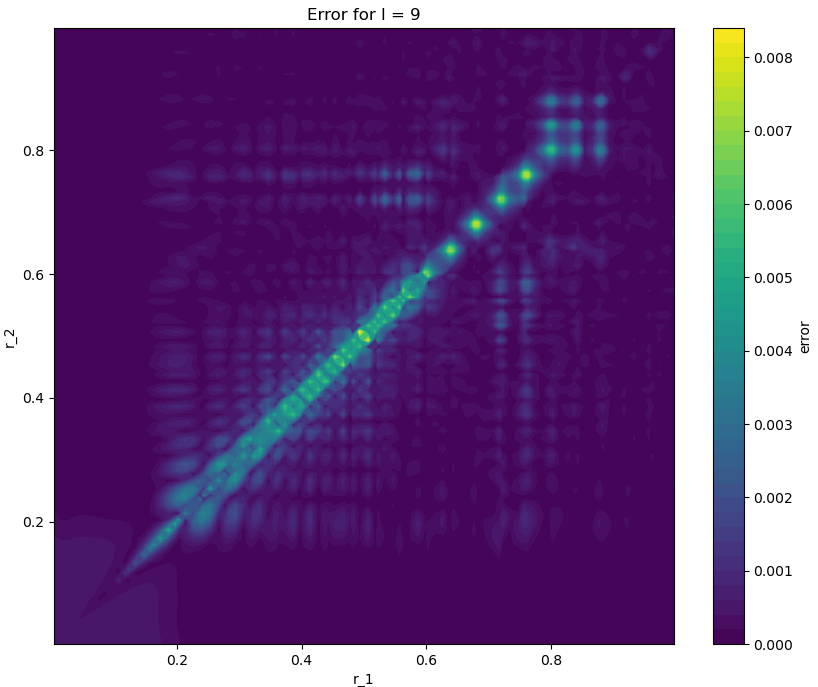}\label{fig:image4}
\end{minipage}    
\caption{Interpolation error for different $\ell$}\label{fig:fourimages}
\end{figure}

The interpolation error comparison for \(\ell = 0, 3, 6, 9 \) is illustrated in 
Figure~\ref{fig:fourimages}, 
revealing that as \( \ell \) increases, the regions of high error become more concentrated 
along the diagonal line \( y = x \). Moreover, the width of the error band around the diagonal 
shows no significant increase, indicating the effectiveness of the adaptive interpolation method.

\subsection{Numerical experiments on helium atom}
In this subsection, we utilize the machine learning method developed 
in Section \ref{Section_Eigenvalue} to determine the wave function by 
solving the associated optimization problem (\ref{loss}). After implementing the  
TNN expansions for discrete Gaussian integration points 
in Subsection \ref{Subsection_TNN_Gauss}, all components can 
be represented using tensor structures. By leveraging the efficiency of 
the TNN framework, multidimensional integrals are transformed 
into products of one-dimensional integrals, 
leading to a significant enhancement in computational efficiency.

Table \ref{He1 error} summarizes the errors associated with various Gaussian quadrature points, 
selected after coordinate transformation, along with different neural network parameters. In our experiments, 
the first part of the parameter column in Table \ref{He1 error} specifies the grid parameters in the 
radial direction (\(r\)). For example, ``$4\times 25$'' indicates that the interval \([0,1]\) is divided into $25$  
subintervals, with four Gaussian nodes placed within each interval. In the \(\theta\) and \(\varphi\) directions, 
the intervals \([0,\pi]\) and \([0,2\pi]\) are divided into 10 and 20 subintervals, 
respectively, with each interval using the same number of Gaussian nodes as specified for the radial direction.

In Table \ref{He1 error}, the parameter \(p\) denotes the rank of the TNN structure. 
The first two rows of the table provide details on the number of layers in the neural network 
along the \(r\), \(\theta\), and \(\varphi\) directions. As previously mentioned, 
we approximate the term \(\int_{D}\frac{\Psi^{2}}{r_{12}}\,dD\) using spherical harmonic expansion. 
Given that this series is infinite, we truncate it to a finite number 
of terms for computing the final ground state energy. 
The parameter \(n\) represents the number of truncation terms used in the spherical harmonic expansion, 
as defined in \eqref{truncation}. The activation function employed is the \(\tanh\) function.

For the helium atom calculations, we utilized the Adam optimizer with a learning rate of \(1 \times 10^{-4}\) 
and trained over 1,500,000 steps. The term \({\rm Error}_{\rm lowest}\) represents the relative error between 
the minimum ground state energy achieved during optimization and the known standard value. 
\({\rm Error}_{\rm best}\) 
denotes the relative error of the ground state energy that is closest to the standard value obtained during optimization. 
\({\rm Error}_{\rm average}\) refers to the average loss values computed over three intervals: 
300,000 steps, 300,000 steps, and 100,000 steps, centered around the epoch with the best test error.

\begin{table}[htbp]
\begin{center}
\caption{Ground state energy error of He}\label{He1 error}
\renewcommand{\arraystretch}{1.5}
\resizebox{\textwidth}{!}{
\begin{tabular}{lcccccc}
\hline
\multicolumn{2}{c}{${\rm FNN}_r$} & ${\rm FNN}_\theta$& \ \ \ 
    ${\rm FNN}_\varphi$ \\ \hline
\multicolumn{2}{c}{(1,30,30,50)} &  (1,20,20,50)& \ \ \ (1,20,20,50) \\ 
\hline
Parameters &${\rm Error}_{\rm lowest}$       &${\rm Error}_{\rm best}$ & ${\rm Error}_{\rm average}$\\ 
\hline
$4\times 25$, $p=50$, $n=20$  &3.356093833168120E-05  &-3.559429939675050E-10 &4.082875081367040E-07\\
$8\times 20$, $p=50$, $n=30$  &1.329360227686580E-05  &2.336020701329850E-09  &4.206106945562550E-06\\
$8\times 20$, $p=50$, $n=20$  &8.938809036191620E-06  &1.714849308980520E-09  &1.164228948335800E-08\\ 
\hline
\end{tabular}}
\end{center}
\end{table}

Note that in the above three training processes with a learning rate of 1e-4, after enough steps, 
the ground state energy will fluctuate around the standard value with an amplitude of 1e-5. 
However, after taking the mean value, it can be seen from the above experiments that our 
calculation results still have a high accuracy.

\begin{figure}[h]
\centering
\begin{minipage}{0.49\textwidth}
\centering
\includegraphics[width=\textwidth]{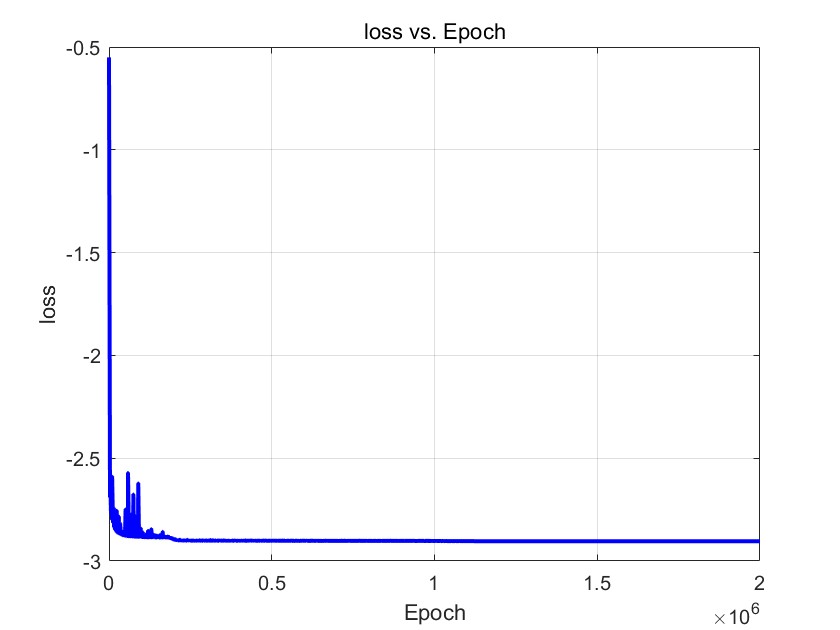}\label{fig:loss}
\end{minipage}
\hspace{0.1pt}
\begin{minipage}{0.49\textwidth}
\centering
\includegraphics[width=\textwidth]{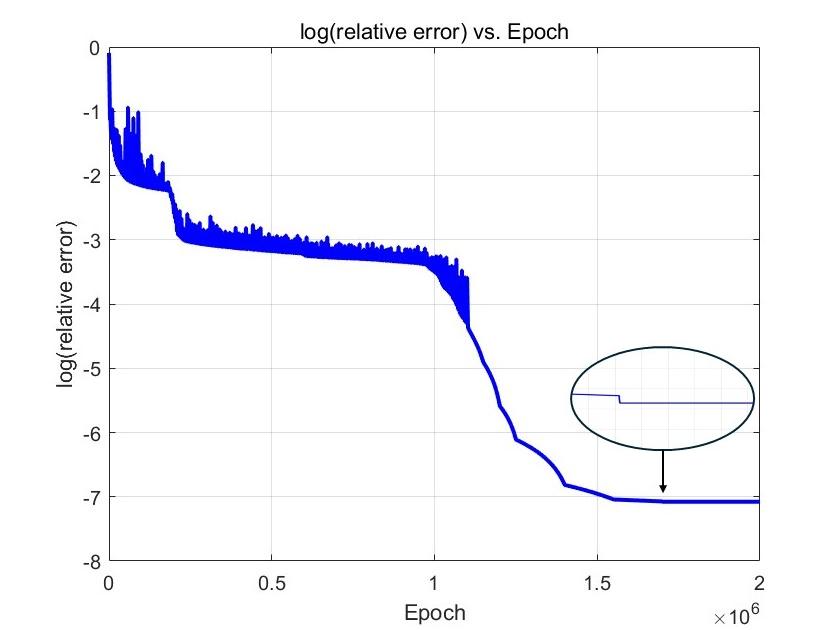}\label{fig:logerr}
\end{minipage}
\caption{Results for He}\label{fig:HE}
\end{figure}

In order to obtain accurate results, 
we adopt the strategy of reducing the learning rate to stabilize the approximate energies. 
The grid parameters in the $r$ direction is $8\times 20$, $p = 50$, $n = 80$. Activation function 
remains the $\tanh$ function. We use the Adam optimizer to train 1,100,000 steps at a learning rate of 1e-04, 
with this number of steps chosen based on the observed oscillations and fluctuations. 
Subsequently, we train for 50,000 steps at learning rates of 1e-6, 1e-7, and 1e-8,  
followed by 150,000 steps at  1e-9, 1e-10, and 1e-11. 
Finally, the LBFGS optimizer is employed for 300,000 steps to achieve the final result. 
In Figure \ref{fig:HE}, we plot the loss and
log(error) as the epoch changes, where error 
is the relative error of approximate energy compared to the standard value. 
As shown in Figure \ref{fig:HE}, the stability of the loss greatly improves 
after decreasing the learning rate, ultimately reaching a convergence value 
with the LBFGS optimizer. The final approximate ground state energy 
is -$\bf{2.903724133801756}$, with a relative error of $\bf{8.37696586e}$-$\bf{08}$ \cite{hiroyuki2007solving}. 

\subsection{Numerical experiments on lithium atom}
For the lithium atom, the Schr\"{o}dinger equation in Cartesian coordinates is given by:
\begin{equation}
-\frac{1}{2}\Delta\Psi - \frac{3\Psi}{r_1} - \frac{3\Psi}{r_2} - \frac{3\Psi}{r_3} + \frac{\Psi}{r_{12}} 
+ \frac{\Psi}{r_{13}} + \frac{\Psi}{r_{23}} = E\Psi,
\end{equation}
where \(\Psi = \Psi(x_1, y_1, z_1, x_2, y_2, z_2, x_3, y_3, z_3)\).


In numerical computations, we truncate the integration domain to a bounded region 
$D \subset \mathbb{R}^9$, which in spherical coordinates is denoted as 
$\widehat{D} = [0,12]^3\times [0,\pi]^3\times[0,2\pi]^3$. As with the helium atom case,  
we perform a coordinate transformation in the $r$ direction, mapping $[0,12]^2$ into $[0,1]^2$.  
Thus for $r_i = s \cdot  t_i$ with $s=12$, we have $\bar{D} = [0,1]^2 \times [0,\pi]^2 \times [0,2 \pi]^2$,  
and the variational principle is
\begin{equation*}
E_1 = \inf_{\Psi}\frac{ \frac{1}{2} \int_{\bar{D}}|\nabla \Psi|^2dx 
-3\int_{\bar{D}}s\left(\frac{1}{r_1}+\frac{1}{r_2}+\frac{1}{r_3}\right)\Psi^2 dx 
+ \int_{\bar{D}}s\left( \frac{1}{r_{12}} + \frac{1}{r_{13}} 
+\frac{1}{r_{23}}\right)\Psi^2dx}{\int_{\bar{D}} s^2 \Psi^2 dx},
\end{equation*}
where $\Psi = \Psi(r_1, \theta_1, \varphi_1, r_2, \theta_2, \varphi_2, r_3, \theta_3, \varphi_3) 
= \Psi(s\cdot  t_1, \theta_1, \varphi_1, s\cdot  t_2, \theta_2, \varphi_2, s\cdot  t_3, \theta_3, \varphi_3)$.

Unlike the helium atom, the anti-symmetry property of lithium is ensured 
by incorporating penalty terms into the loss function. Thus we use the following loss function 
\begin{equation}
L[\Psi] = \frac{\langle \Psi|\widehat H|\Psi\rangle}{\langle \Psi|\Psi\rangle}
+k \cdot  \frac{\langle T_{12}\Psi|\Psi\rangle }{ \langle\Psi|\Psi\rangle },
\end{equation}
where
\begin{eqnarray}
\langle T_{12}\Psi|\Psi\rangle &=& \sum_{j=1}^{p}\sum_{i=1}^{p} \int_0^{12} r_1^2\phi_{r_1, i}(r_1)
\phi_{r_2, j}(r_1)dr_1 \int_0^{12} r_2^2\phi_{r_1, i}(r_2)\phi_{r_2, j}(r_2)dr_2\nonumber\\
&&\int_0^{12} r_3^2\phi_{r_3, i}(r_3)\phi_{r_3, j}(r_3)dr_3
\int_0^{\pi} \sin{\theta_1} \phi_{\theta_1, i}(\theta_1)\phi_{\theta_2, j}(\theta_1)d\theta_1 \nonumber\\
&&\int_0^{\pi} \sin{\theta_2} \phi_{\theta_1, i}(\theta_2)\phi_{\theta_2, j}(\theta_2)d\theta_2 
\int_0^{\pi} \sin{\theta_3} \phi_{\theta_3, i}(\theta_3)\phi_{\theta_3, j}(\theta_3)d\theta_3\nonumber\\
&&\int_0^{2\pi} \phi_{\varphi_1, i}(\varphi_1)\phi_{\varphi_2,j}(\varphi_1)d\varphi_1
\int_0^{2\pi} \phi_{\varphi_1, i}(\varphi_2)\phi_{\varphi_2,j}(\varphi_2)d\varphi_2\nonumber\\
&&\int_0^{2\pi} \phi_{\varphi_3, i}(\varphi_3)\phi_{\varphi_3,j}(\varphi_3)d\varphi_3.
\end{eqnarray}

In our experiment, we initially choose the penalty parameter as $k = 100$. 
For the lithium atom calculation, we utilized the Adam optimizer with a learning rate of 1e-4 
and trained for 300,000 steps. During the training process, the anti-symmetry of the 
lithium atom is enforced by adding a penalty term to the loss function, 
which introduce some instability due to optimization challenges. 
To achieve more accurate results, we adopt a strategy of reducing the learning rate and adjusting 
the hyperparameter $k$ to 10,000. 
Although this requires more time, it significantly improves the precision of the anti-symmetry term, 
enhancing the stability of the calculations.
The grid parameters in the $r$ direction are set to $8 \times 25$, 
with the rank parameter of TNN being $p = 50$ and the truncation number $n = 140$. 
The activation function remains the $\tanh$ function. Using the Adam optimizer, 
we trained for 1,050,000 steps at a learning rate of 1e-06, followed by 250,000 steps at 
1e-08, 150,000 steps at 1e-10, and finally 500,000 stepsat 1e-12.

\begin{figure}[h]
\centering
\begin{minipage}{0.45\textwidth}
\centering
\includegraphics[width=\textwidth]{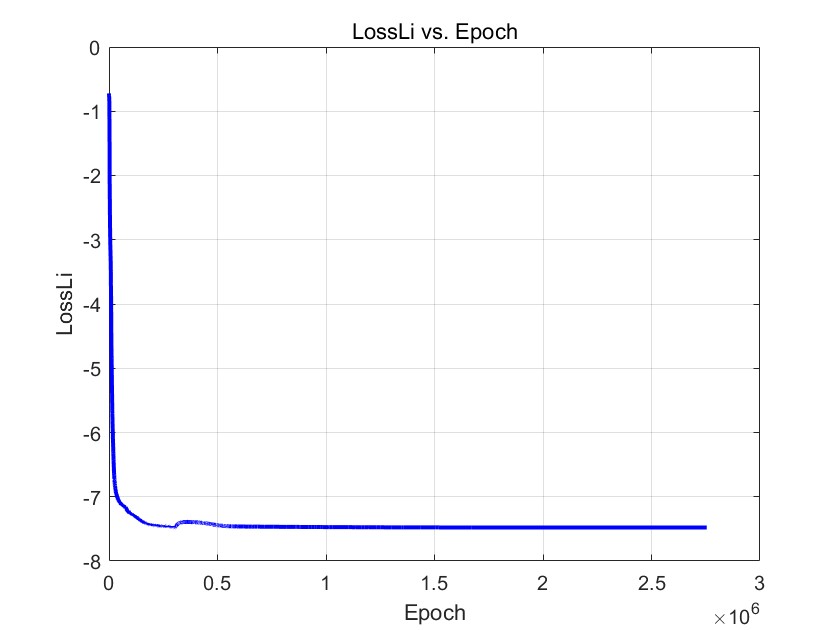}
\label{fig:loss}
\end{minipage}\hspace{0.1pt}
\begin{minipage}{0.45\textwidth}
\centering
\includegraphics[width=\textwidth]{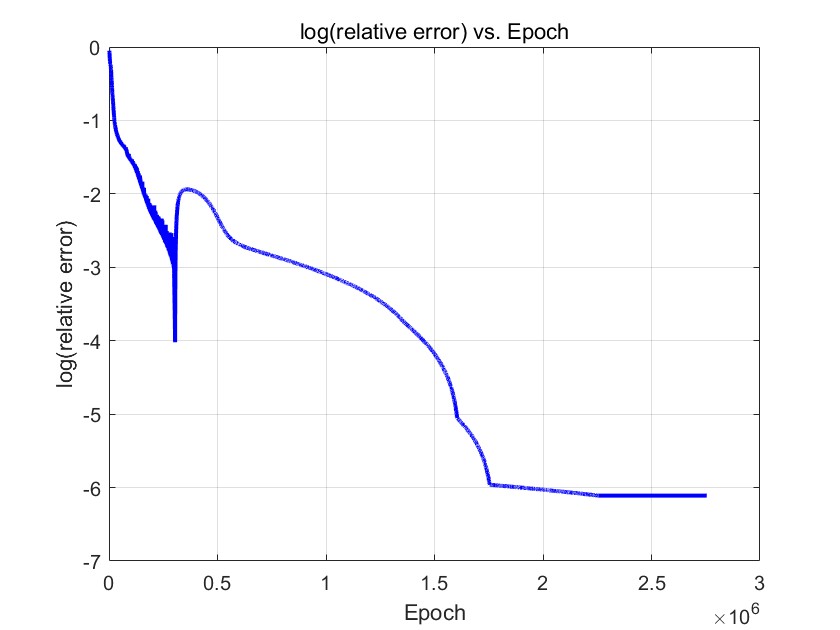}
\label{fig:logerr}
\end{minipage}
\caption{Results for Li}
\label{fig:LI}
\end{figure}

Subsequently, we use the LBFGS optimizer for additional 300,000 steps to obtain the final result.  
In Figure \ref{fig:LI}, we plot the loss and log(error) against the number of epochs, 
where error represents the relative error 
of the approximate energy compared to the standard value. 
From Figure \ref{fig:LI}, we can see that although 
the ground state energy loss increased slightly after changing the hyperparameter $k$, 
the stability of the loss improves significantly after reducing the learning rate. 
The loss eventually converges, maintaining high precision for the 
anti-symmetry term after training with LBFGS. 
The final approximate ground state energy is -$\bf{7.4780545032264572}$, 
with a relative error of $\bf{7.77846092e}$-$\bf{07}$ \cite{david2024accurate}.

\subsection{Numerical experiments on hydrogen molecule}
The hydrogen molecule (${\rm H}_2$) consists of two positively charged hydrogen atoms, 
each contributing one electron to form a shared electron pair. The Hamiltonian operator 
$\widehat{H}$ for the hydrogen molecule can be expressed in the following form
\begin{equation}\label{Hamilton_H2}
\widehat{H} = -\frac{1}{2} \sum_{i=1}^{2} \Delta_i -
\frac{1}{2} \sum_{j=1}^{2} \Delta_j - \frac{1}{r_{12}} 
- \sum_{j=1}^{2}\left(\frac{1}{r_{j1}} + \frac{1}{r_{j2}}\right) + \frac{1}{r_{ab}},
\end{equation}
where $\Delta_i$ and $\Delta_j$ are the Laplacian operators for the electrons 
and nuclei, respectively; $r_{12}$ is the distance between the two electrons;  
$r_{j1}$ and $r_{j2}$ are the distances between the electrons and the nuclei;  
and $r_{ab}$ is the distance between the two nuclei. In (\ref{Hamilton_H2}), 
the first term represents the kinetic energy of the electrons, 
the second term represents the kinetic energy of the nuclei, 
the third term denotes the electron-electron repulsion, 
the fourth and fifth terms account for the electron-nucleus attractions, 
and the last term represents the nucleus-nucleus repulsion,  
with \( r_{ab} \) fixed at \( 1.4 \, \text{bohr} \).


Similar to the calculations for helium and lithium atoms described above, 
we can also formulate the hydrogen molecule as an optimization problem aimed at 
minimizing the loss function using TNNs. 
In the optimization process for the hydrogen molecule, we select \(8 \times 20\) 
Gaussian nodes in the \(r\) direction, \(8 \times 30\) nodes in the \(\theta\) direction, 
and \(8 \times 60\) nodes in the \(\psi\) direction. 
The sizes of TNN are set to \([1, 30, 30, 50]\) and \([1, 20, 20, 50]\), respectively. 
The spherical harmonics expansion is truncated at \(n = 60\). 
Due to observed oscillations with the Adam optimizer during the optimization process, 
we employ a progressively decreasing learning rate approach to achieve higher precision, 
with final convergence performed using the LBFGS optimizer. 
Unlike the helium and lithium atoms, the optimization of the hydrogen 
molecule is significantly affected by a large proportion of singular terms. 
As a result, the Galerkin method-based loss function optimization is very slow. 
To address this, we introduce the Ritz form of the least-squares method. 
From a mathematical point of view, this can be understood as first using 
neural network optimization to identify a larger subspace, 
followed by applying the least-squares method to search 
for the extremum within this initialized subspace.

\begin{figure}[h]
\centering
\begin{minipage}{0.45\textwidth}
\centering
\includegraphics[width=\textwidth]{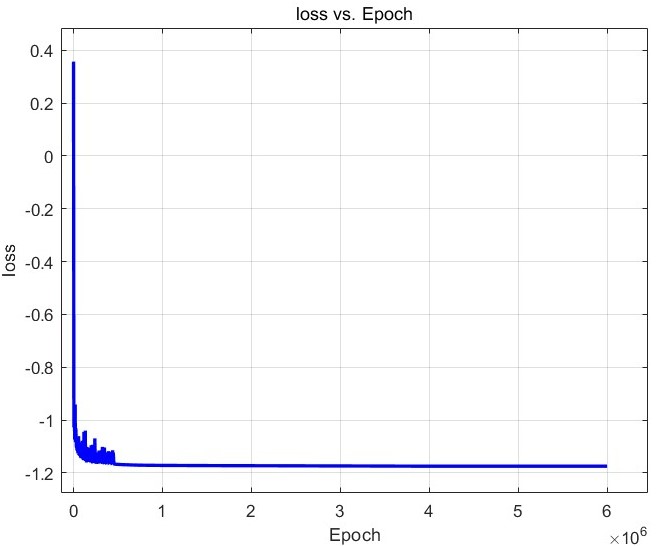}
\label{fig:loss}
\end{minipage}\hspace{0.1pt}
\begin{minipage}{0.45\textwidth}
\centering
\includegraphics[width=\textwidth]{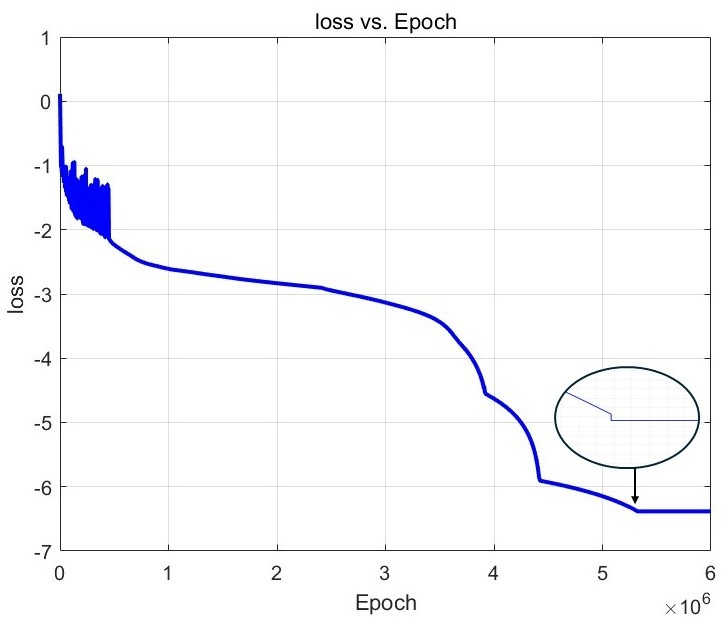}
\label{fig:logerr}
\end{minipage}
\caption{Result for ${\rm H}_2$}\label{fig:H2}
\end{figure}

Using the Adam optimizer, we trained for 447,500 steps at a learning rate of \(1 \times 10^{-4}\), 
with this number of steps chosen based on observed oscillations and fluctuations. 
We then trained for 650,000 steps at a learning rate of \(1 \times 10^{-6}\), 
followed by 1,400,000 steps at \(1 \times 10^{-7}\) using the Galerkin method, 
2,900,000 steps at \(1 \times 10^{-7}\) using the Ritz method, 
500,000 steps at \(1 \times 10^{-9}\) using the Ritz method, 
and finally 900,000 steps at \(1 \times 10^{-11}\) using the Ritz method. 
The LBFGS optimizer was then employed to obtain the final result. 
In Figure \ref{fig:H2}, we plot the loss and \(\log(\text{error})\) as the epochs changing, 
where error is the relative error of the approximate energy  
compared to the standard value. 
As shown in Figure \ref{fig:H2}, the stability of the loss significantly is 
improved after decreasing the learning rate, ultimately reaching a convergence 
value after using LBFGS. The final approximate ground state energy 
is -\(\mathbf{1.1744752290871725}\), with a relative error of 
\(\mathbf{4.13063311e}$-$\bf{7}\) \cite{james2006high}.

\section{Conclusions}
The aim of this paper is to propose a machine learning method to address one of 
the most significant high-dimensional problem: the Schrödinger equation. 
The main contribution of this work is the first-time application 
of the TNN structure to solve the Schr\"{o}dinger equation, 
along with the corresponding machine learning methodology.

Based on the high accuracy and high efficiency quadrature scheme 
for the TNN structure, we design the TNN interpolation and TNN expansion methods 
for Coulomb potential terms. 
Additionally, we incorporate anti-symmetric penalty terms in the form of 
the inner product to satisfy Pauli exclusion principle.
Experimental results demonstrate that the proposed TNN-based machine learning method 
can effectively solve  the ground state of many-electron systems 
while maintaining manageable computational complexity. 
Furthermore, we report preliminary work on computing the ground state molecular 
structures of two diatomic molecules to further illustrate the efficacy of 
TNN in solving Schr\"{o}dinger equations, providing a novel perspective 
for bond length calculations. 

It is noteworthy that, due to the high accuracy of high-dimensional 
integration of TNN functions, 
the multiple energy levels of small particle systems, such as helium atoms, 
lithium atoms, and hydrogen molecules, can be readily computed \cite{WangXie}. 

Based on the analysis and numerical experiments presented in this paper, 
we believe that the TNN method holds significant potential for solving more complex Schr\"{o}dinger equations. 
In future work, a broader range of electronic systems should be considered to evaluate 
the performance of TNN on Schr\"{o}dinger equations. 
Additionally, some methodological details require further refinement.
First, we should explore the application of the Slater determinant method to address 
the anti-symmetry property, aiming for more stable and accurate results.
Second, we will also consider the excited states of Schr\"{o}dinger equations, 
utilizing a similar approach in \cite{WangXie}.

\end{document}